\documentclass[12pt]{article}
\pdfoutput=1
\usepackage{jheppub}

\usepackage{amsmath,bbm,array,amsfonts,graphicx,wrapfig,lscape,float,mathtools,multirow,longtable}

\newcommand{\be}{\begin{equation}}
\newcommand{\ee}{\end{equation}}
\newcommand{\beq}{\begin{equation}}
\newcommand{\beql}[1]{\begin{equation}\label{#1}}
\newcommand{\eeq}{\end{equation}}
\newcommand{\ba}{\begin{array}}
\newcommand{\ea}{\end{array}}
\newcommand{\bea}{\begin{eqnarray}}
\newcommand{\beal}[1]{\begin{eqnarray}\label{#1}}
\newcommand{\eea}{\end{eqnarray}}
\newcommand{\ben}{\begin{enumerate}}
\newcommand{\een}{\end{enumerate}}
\newcommand{\bean}{\begin{eqnarray*}}
\newcommand{\eean}{\end{eqnarray*}}
\newcommand{\eref}[1]{(\ref{#1})}
\newcommand{\sref}[1]{\S\ref{#1}}
\newcommand{\tref}[1]{Table~\ref{#1}}
\newcommand{\nn}{\nonumber}

\newcommand{\fref}[1]{Figure \ref{#1}}
\newcommand{\btab}[1]{\begin{tabular}{#1}}
\newcommand{\etab}{\end{tabular}}
\newcommand{\diag}{\mbox{diag}}

\newcommand{\comment}[1]{}

\newcommand{\ud}{\mathrm{d}}

\newcommand{\PE}{\mathrm{PE}}
\newcommand{\PL}{\mathrm{PL}}

\newcommand{\qed}{\nobreak \ifvmode \relax \else
      \ifdim\lastskip<1.5em \hskip-\lastskip
      \hskip1.5em plus0em minus0.5em \fi \nobreak
      \vrule height0.75em width0.5em depth0.25em\fi}

\title{Hilbert Series for Theories with Aharony Duals}
\author[a]{Amihay Hanany,}
\author[b]{Chiung Hwang,}
\author[b]{Hyungchul Kim,}
\author[b,c]{Jaemo Park,}
\author[d]{and Rak-Kyeong Seong}

\affiliation[a]{
Theoretical Physics Group, Blackett Laboratory, \\
Imperial College London, Prince Consort Road,
London SW7 2AZ, United Kingdom
}

\affiliation[b]{
Department of Physics, POSTECH, \\
Pohang 790-784, Korea
}

\affiliation[c]{
Postech Center for Theoretical Physics (PCTP), POSTECH,\\
Pohang 790-784, Korea
}

\affiliation[d]{
School of Physics, Korea Institute for Advanced Study, \\
85 Hoegi-ro, Seoul 130-722, Korea
}

\emailAdd{a.hanany@imperial.ac.uk}
\emailAdd{c\_hwang@postech.ac.kr}
\emailAdd{dakiro@postech.ac.kr}
\emailAdd{jaemo@postech.ac.kr}
\emailAdd{rkseong@kias.re.kr}

\preprint{
\begin{flushright}
KIAS-P14084
\\
IMPERIAL-TP-15-AH-01
\end{flushright}
}

\abstract{
The algebraic structure of moduli spaces of 3d $\mathcal{N}=2$ supersymmetric gauge theories is studied by computing the Hilbert series  which is a generating function that counts gauge invariant operators in the chiral ring. These $U(N_c)$ theories with $N_f$ flavors have Aharony duals and their moduli spaces receive contributions from both mesonic and monopole operators. In order to compute the Hilbert series, recently developed techniques for Coulomb branch Hilbert series in 3d $\mathcal{N}=4$ are extended to 3d $\mathcal{N}=2$. The Hilbert series computation leads to a general expression of the algebraic variety which represents the moduli space of the $U(N_c)$ theory with $N_f$ flavors and its Aharony dual theory. A detailed analysis of the moduli space is given, including an analysis of the various components of the moduli space.
}

\begin{document}

\maketitle

%==================================================%
%==================================================%
\section{Introduction}

Dualities between supersymmetric gauge theories have attracted much interest in the past. In particular, dualities have shed light on understanding the strongly coupled regime of supersymmetric gauge theories. One way to identify dual supersymmetric gauge theories is to understand the structure of their vacuum moduli spaces. Recently, tools such as the Hilbert series \cite{Benvenuti:2006qr,Hanany:2006uc,Feng:2007ur,Butti:2007jv,Hanany:2007zz,Forcella:2008bb,Forcella:2008eh,Cremonesi:2013lqa} have been effectively used to obtain a better understanding of vacuum moduli spaces of various supersymmetric gauge theories.

Seiberg duality \cite{Seiberg:1994pq}, proposed 20 years ago, is a quintessential example of an IR duality that relates $\mathcal{N}=1$ SQCD theories with gauge group $SU(N_c)$ and $N_f$ flavors with $SU(N_f-N_c)$ gauge theories with $N_f$ flavors. A $3d$ $\mathcal{N}=2$ analog of Seiberg duality was proposed in 1997 \cite{Aharony:1997gp,Aharony:1997bx,Karch:1997ux}. The duality which is now known as Aharony duality relates a $U(N_c)$ theory with $N_f$ chiral fundamental and $N_f$ chiral anti-fundamental multiplets with a \textit{dual} $U(N_f-N_c)$ theory with $N_f$ chiral fundamentals and $N_f$ chiral anti-fundamentals. These Aharony dual theories have been studied extensively in the past, with attempts to match the chiral rings of dual theories, in particular by computing the corresponding superconformal indices \cite{Witten:1999ds,Kapustin:2011vz,Kim:2009wb,Imamura:2011su,Bhattacharya:2008bja,Bhattacharya:2008zy,Hwang:2011ht,Bashkirov:2011vy,Hwang:2011qt,Kim:2013cma}. In this work, we want to express the moduli space of Aharony dual theories as an affine algebraic variety by computing the Hilbert series.

Hilbert series are generating functions which count gauge invariant operators in the chiral ring of the supersymmetric gauge theory. They have been used to extract information about the exact algebraic structure of vacuum moduli spaces \cite{Benvenuti:2006qr,Hanany:2006uc,Feng:2007ur}. For instance, Hilbert series for instanton moduli spaces \cite{Benvenuti:2010pq,Hanany:2012dm,Dey:2013fea,Cremonesi:2014xha} and vortex moduli spaces \cite{Hanany:2014hia} have shed light on the algebraic structure of the corresponding moduli spaces. Moreover, $4d$ $\mathcal{N}=1$ theories represented by bipartite graphs on the torus known as brane tilings \cite{Hanany:2012hi,Hanany:2012vc} have been studied with the help of Hilbert series. More recently, techniques have been developed for computing the Hilbert series for the Coulomb branch moduli space of $3d$ $\mathcal{N}=4$ theories in \cite{Cremonesi:2013lqa,Cremonesi:2014kwa,Cremonesi:2014vla} and \cite{Cremonesi:2014xha} which paved the way in further understanding among other things instanton moduli spaces as Coulomb branches of extended Dynkin diagrams.

In this work, we want to express the moduli space of $3d$ $\mathcal{N}=2$ Aharony dual theories as an algebraic variety. In order to compute the Hilbert series, recently developed techniques for Coulomb branch Hilbert series in $3d$ $\mathcal{N} = 4$ \cite{Cremonesi:2013lqa} are extended to $3d$ $\mathcal{N} = 2$. Given the Hilbert series, it is possible using plethystics \cite{Benvenuti:2006qr,Feng:2007ur,Hanany:2007zz} to extract information about the generators and first order relations amongst the generators of the moduli space.

The moduli space for $3d$ $\mathcal{N}=2$ supersymmetric gauge theories is the space of dressed monopole operators. These operators are dressed with gauge invariant operators which are invariant under a residual gauge symmetry left unbroken under the monopole background. Furthermore, the moduli space is partially lifted due to instanton effects \cite{Aharony:1997gp,Aharony:1997bx,Karch:1997ux,Aharony:2013dha,callias1978}. As such, methods for the Coulomb branch Hilbert series for $3d$ $\mathcal{N}=4$ theories can be generalized for Aharony dual theories. In this work, we use a  sum over a sublattice of GNO charges for the monopole operators which are \textit{dressed} by suitable gauge invariant operators. The sum over the GNO sublattice generates the Hilbert series of the moduli space. By doing so, we are able to express the moduli space as an algebraic variety for any $U(N_c)$ gauge theory with $N_f$ flavors and their Aharony dual theory.

Our Hilbert series computation identifies the generators of the moduli space which agree with previously known results \cite{Bashkirov:2011vy}. Moreover, since the Hilbert series computation gives the algebraic structure of the chiral ring, including relations amongst the generators,\footnote{This is up to numerical coefficients which can usually be absorbed into the elements of the chiral ring. In this work, the numerical coefficients are not needed as the relations are homogeneous and there is precisely one operator per relation.} we are able to study in detail the structure of the vacuum moduli space, including the structure of its components.

This work compares the Hilbert series with the superconformal index for Aharony dual theories. It is important to note that in order to know the entire algebraic structure of the moduli space, it is crucial to compute the Hilbert series directly. The superconformal index gives information on the moduli space only after one finds an appropriate limit to a Hilbert series. 

The work is structured as follows: section \sref{s1} introduces the $3d$ $\mathcal{N}=2$ supersymmetric gauge theories which are discussed in this paper. Section \sref{s2} introduces the Hilbert series and the method used to compute it for these theories. In particular, the section outlines the structure of the partially lifted GNO charge lattice and the summation of the dressed monopole operators which is necessary for the computation of the Hilbert series. Explicit examples up to gauge group $U(3)$ are given and a generalization of the algebraic variety for the moduli space is presented. A further analysis on the various components of the moduli space is presented. Section \sref{scfi} compares the superconformal index with the Hilbert series.
\\

\noindent\textit{Note added:} We acknowledge a future paper to appear in \cite{unpubcremonesi} that also discusses moduli spaces of dressed monopole operators for $3d$ $\mathcal{N}=2$ theories.
\\

%==================================================%
%==================================================%
\section{The Theory and Aharony Duality \label{s1}}

\begin{figure}[ht!!]
\begin{center}
\includegraphics[trim=0cm 0cm 0cm 0cm,width=0.7\textwidth]{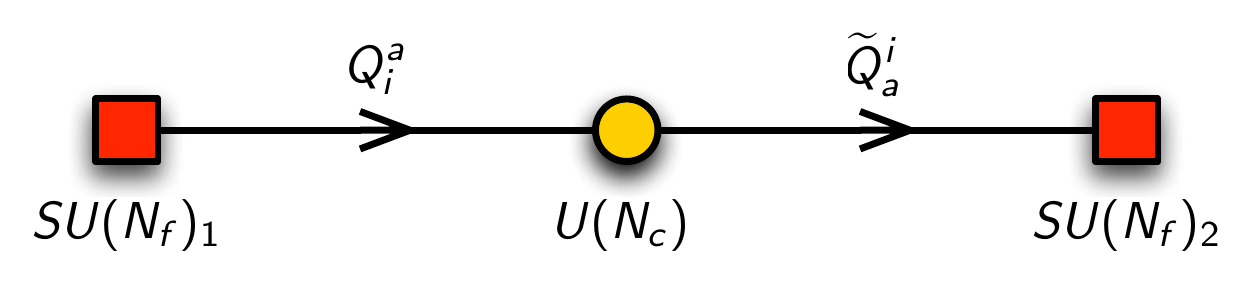}
\caption{
The quiver diagram for the 3d $\mathcal{N}=2$ theory with a $U(N_c)$ gauge group and $N_f$ flavors. 
\label{fquiver}
}
 \end{center}
 \end{figure}

\paragraph{The Theory.}
We are interested in the moduli space of a 3d $\mathcal{N}=2$ $U(N_c)$ gauge theory with $N_f$ flavors that has a global symmetry $S(U(N_f)_1 \times U(N_f)_2)$. The vector multiplet of the theory contains the adjoint real scalar $\sigma$ and the gauge field $A$. The scalar can be diagonalised to give $\sigma=\diag(\sigma_1,\dots,\sigma_{N_c})$. The theory also has chiral multiplets containing chiral matter fields $Q$ and $\tilde{Q}$ which respectively transform in the fundamental and anti-fundamental representations of the gauge group $U(N_c)$. The corresponding quiver diagram of the theory is shown in \fref{fquiver}.

\begin{figure}[ht!!]
\begin{center}
\includegraphics[trim=0cm 0cm 0cm 0cm,width=0.6\textwidth]{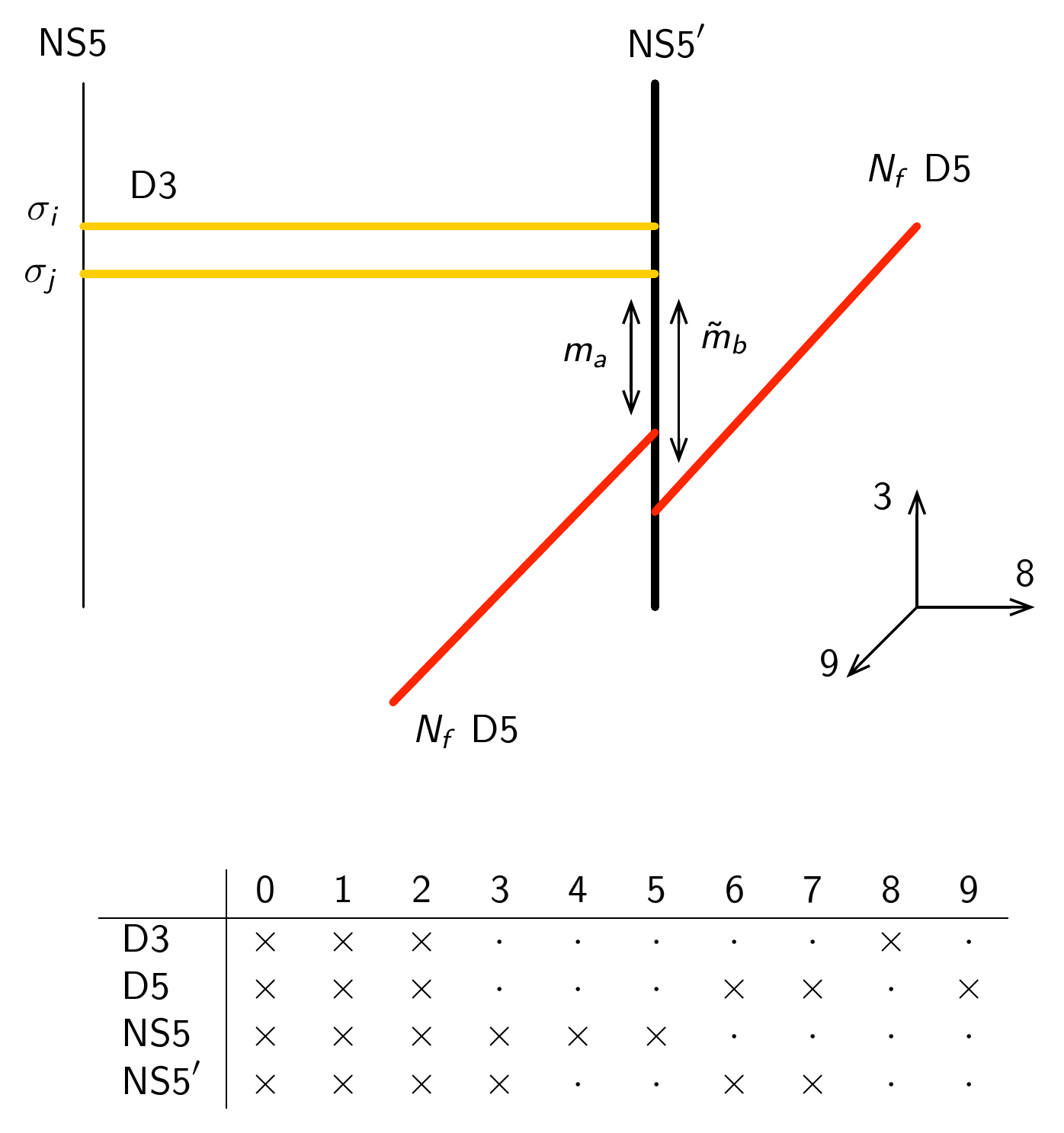}
\caption{
The brane construction for the $3d$ $\mathcal{N}=2$ $U(N_c)$ theory with $N_f$ flavors that has a global symmetry $S(U(N_f)_1 \times U(N_f)_2)$. There are $N_c$ D3-branes suspended between 2 NS5-branes. The positions of the branes along the $x^3$-direction are given by the scalar adjoints $\sigma_i$, where $i=1,\dots,N_c$. There are also $N_f$ pairs of D5-branes which are attached to the $\text{NS5}^\prime$-branes along the $x^9$-direction. The position along the $x^3$-direction for the D5-branes are given by the real masses $m_a$ and $\tilde{m}_b$ for $Q$ and $\tilde{Q}$ respectively, where $a,b=1,\dots,N_f$.
\label{fbrane}
}
 \end{center}
 \end{figure}

The theory can be realized with D3 branes in a D5 and NS5-brane background \cite{Aharony:1997ju} as shown in \fref{fbrane}. The $N_c$ D3-branes are suspended between 2 NS5-branes and their positions along the $x^3$-direction are labelled by $\sigma_i$, where $i=1,\dots,N_c$. For each of the flavour groups $U(N_f)_1$ and $U(N_f)_2$, there is a stack of $N_f$ D5 branes attached to the $\text{NS5}^\prime$ along the $x^9$-direction. Their positions along the $x^3$-direction are respectively labelled by the real masses $m_a$ and $\tilde{m}_b$ of $Q$ and $\tilde{Q}$ where $a,b=1,\dots, N_f$. For the theories considered here, the bare masses are set to zero.
\\

The moduli space of the $3d$ $\mathcal{N}=2$ $U(N_c)$ theory receives quantum corrections. The Higgs branch is parameterized by mesonic operators of the form $M=Q\tilde{Q}$ which are invariant under the gauge group $U(N_c)$. The remaining moduli space is parameterized by chiral operators that are composed of supersymmetrized 't Hooft monopole operators $v_{m}$ with magnetic charge $m$ and mesonic operators of the form $M_{m}=Q\tilde{Q}$ which are invariant under a residual subgroup $H_{m} \subset U(N_c)$. In other words, there are chiral gauge invariant operators which are either \textit{bare} monopole operators built out of $v_{m}$, or \textit{dressed} monopole operators which are built out of the mesonic operators $M_m$ and bare monopole operators $v_{m}$. 

The 't Hooft monopole operators are defined by introducing a Dirac monopole singularity at an insertion point in the Euclidean path integral \cite{'tHooft19781}. By Dirac quantization, the monopole operators are labelled by magnetic charges on a weight lattice $\Gamma_{G^\vee}$ of the GNO/Langlands dual group $G^{\vee}$ \cite{PhysRevD.14.2728,Goddard19771,Kapustin:2005py}. For gauge group $G=U(N_c)$, the magnetic charge takes the form
\beal{es1000a1}
m= (m_1,m_2, \dots , m_{N_c}) ~,~
\eea
where by fixing the action of the Weyl symmetry $W_{G}$ $m_1 \geq m_2 \geq \dots \geq m_{N_c}$ such that $m\in \Gamma_{G^\vee}/W_{G}$. Note that the magnetic charges $m_i$ can be considered conjugate to the $\sigma_i$ of the diagonalised scalar adjoint in the vector multiplet of the theory. 

Instanton effects \cite{Aharony:1997gp,Aharony:1997bx,Karch:1997ux,Aharony:2013dha} lift most of the moduli space of the theory such that magnetic charges of the remaining monopole operators have 
\beal{es1000a2}
m_2 = \dots = m_{N_c - 1 } = 0 ~.~
\eea
The remaining GNO charges are $m_1 \geq 0  \geq m_{N_c}$. For convenience, the index for the magnetic charge variable $m_{N_c}$ is relabelled to $m_2$ such that the magnetic charges of monopole operators are of the form
\beal{es1000a3}
m_1\geq 0 \geq m_{2}
~.~
\eea
We introduce the following notation for bare monopole operators with magnetic charges
\beal{es1000a3}
(m_1,m_2)=(+1,0) &~:~& v_m \equiv  v_{+} ~,~
\nn\\
(m_1,m_2)=(0,-1) &~:~& v_m \equiv  v_{-} ~.~
\eea

The bare and dressed monopole operators have magnetic charges $m_1 \geq 0 \geq m_2$. Non-zero magnetic charges $m_1,m_2$ give effective masses $|\sigma_i - \sigma_j|$ and $|\sigma_i|$ to the gauge field $A$ and the matter fields $Q, \tilde{Q}$ respectively. These massive fields are integrated out with the gauge group $G$ breaking into a residual subgroup $H_m \subset G$. For our theory with gauge group $U(N_c)$, the residual subgroup is one of the following:
\begin{itemize}
\item $m_1,m_2=0$: $U(N_c)$
\item $m_1\neq0, m_2=0$ or $m_1 =0, m_2\neq0$: $U(N_c-1) \times U(1)$
\item $m_1,m_2\neq0$: $U(N_c-2) \times U(1) \times U(1)$
\end{itemize}  
The above values for $m_1,m_2$ can be thought of as 4 sublattices of the GNO lattice where in each particular sublattice the gauge group $U(N_c)$ breaks into a particular residual subgroup $H_m$.

The global symmetry of our theory is $SU(N_f)_{1}\times SU(N_f)_2 \times U(1)_A \times U(1)_T \times U(1)_R$, where $SU(N_f)_1 \times SU(N_f)_2$ is the flavour symmetry, $U(1)_A$ is the axial symmetry, $U(1)_T$ is the topological symmetry and $U(1)_R$ is the R-symmetry. The global charges carried by the bare monopole operators and matter fields are summarized in \tref{tfields}. 
\\

\begin{table}[ht!]
\centering
\resizebox{1\hsize}{!}{
\begin{tabular}{|c|cc|cc|cc|c|}
\hline
\; & \multicolumn{2}{c|}{$U(N_c)$} & & & & &
\\
\; & $SU(N_c)$ & $U(1)_B$ & $SU(N_f)_1$ & $SU(N_f)_2$ & $U(1)_A$ & $U(1)_T$ & $U(1)_R$
\\
\hline\hline
$Q_{i}^{a}$ & $[1,0,\dots,0]_{z}$ & $+1$ & $[0,\dots,0,1]_{{}u}$ & 0 & $1$ & $0$ & $r$
\\
$\widetilde{Q}_{a}^{i}$ & $[0,\dots,0,1]_{z}$ & $-1$ & 0 & $[1,0,\dots,0]_{\tilde{{}u}}$ & $1$ & $0$ & $r$
\\
$v_{\pm}$ & 0 & 0 & 0 & 0 & $-N_f$ & $\pm 1$ & $(1-r)N_f - (N_c -1)$
\\
\hline
\end{tabular}
}
\caption{
The $U(N_c)$ theory with $N_f$ flavors (theory A). The table shows the fundamental and anti-fundamental matter fields and bare monopole operators under gauge and global symmetries.
\label{tfields}
}
\end{table}

\begin{figure}[ht!!]
\begin{center}
\includegraphics[trim=0cm 0cm 0cm 0cm,width=1\textwidth]{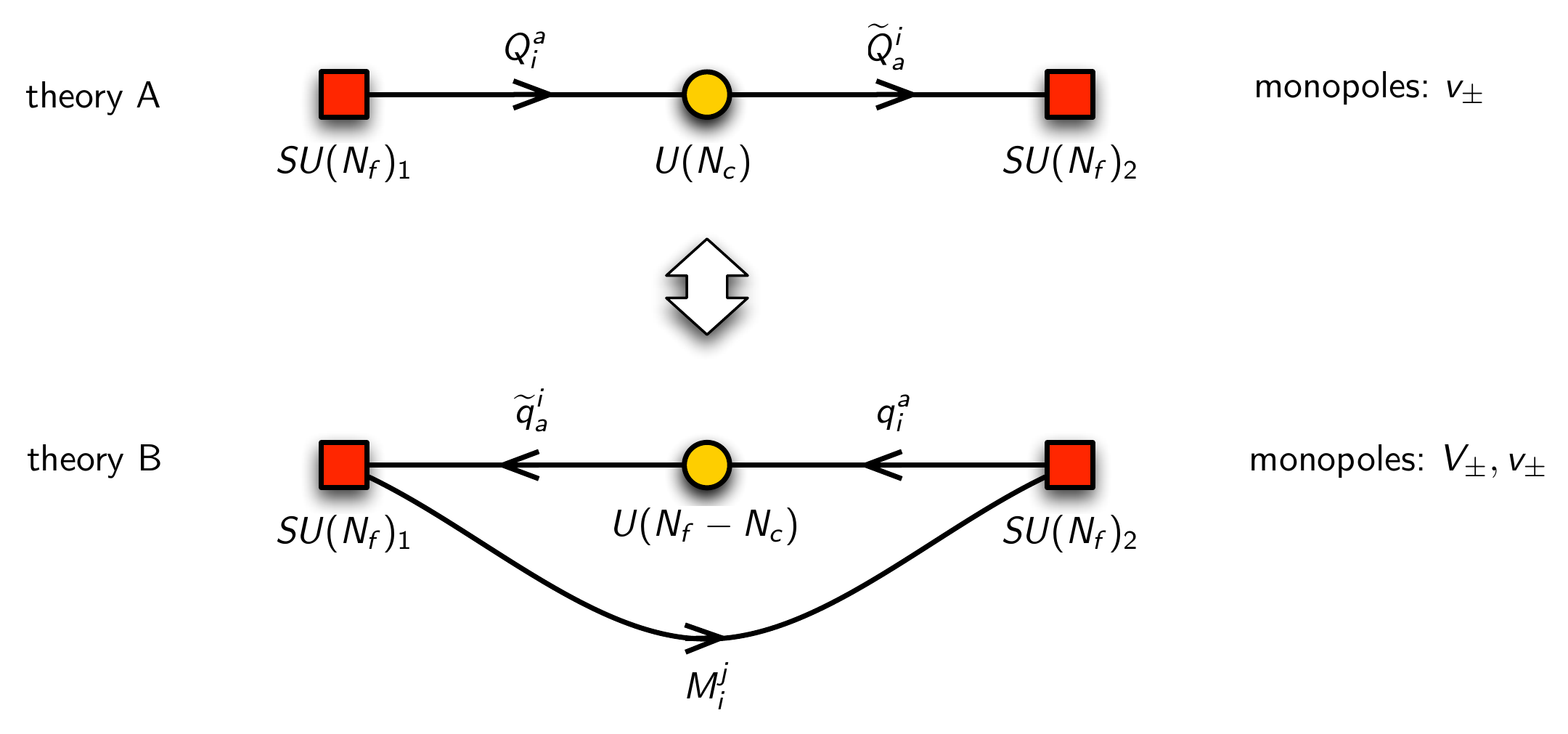}
\caption{
The quiver diagrams of theories A and B under Aharony duality.
\label{fdualquivers}
}
 \end{center}
 \end{figure}

\paragraph{Aharony Duality.} We call the 3d $\mathcal{N}=2$ theory with $U(N_c)$ gauge group and $N_f$ flavors as the \textit{theory A}. \textit{Aharony duality} \cite{Aharony:1997gp,Aharony:1997bx,Karch:1997ux} maps theory A to a new theory for $N_f > N_c$. This dual theory is a $\mathcal{N}=2$ $3d$ theory with $U(N_f-N_c)$ gauge symmetry and $N_f$ flavors and is called \textit{theory B}. The dual theory has $v_{\pm}$ and $M_{i}^{j}$ of theory A as gauge singlets. In addition, there are fundamental $q_{i}^{a}$ and anti-fundamental $\widetilde{q}_{ai}$ under the $U(N_f-N_c)$ gauge symmetry as well as monopole operators $V_{+},V_{-}$ under the gauge group $U(N_f-N_c)$. The monopole operators $V_{\pm}$ respectively carry magnetic charges of the form
\beal{es0a50}
(\pm 1,\underbrace{0,\dots,0}_{N_f-N_c-1})
~.~
\eea
The global symmetry of the theory B is the same as for theory A,  $SU(N_f)_1\times SU(N_f)_2 \times U(1)_A \times U(1)_T \times U(1)_R$. The quiver diagrams of theories A and B are shown in \fref{fdualquivers}. The matter fields and monopole operators under gauge and global symmetries of theory B are summarized in \tref{tfields2}.

\begin{table}[ht!]
\centering
\resizebox{1\hsize}{!}{
\begin{tabular}{|c|cc|cc|cc|c|}
\hline
\; & \multicolumn{2}{c|}{$U(N_f-N_c)$} & & & & &
\\
\; & $SU(N_f-N_c)$ & $U(1)_B$ & $SU(N_f)_1$ & $SU(N_f)_2$ & $U(1)_A$ & $U(1)_T$ & $U(1)_R$
\\
\hline\hline
$q_{i}^{a}$ & $[1,0,\dots,0]_{z}$ & $+1$ & 0 &$[0,\dots,0,1]_{\tilde{{}u}}$ &  $-1$ & $0$ & $1-r$
\\
$\widetilde{q}_{a}^{i}$ & $[0,\dots,0,1]_{z}$ & $-1$ & $[1,0,\dots,0]_{{}u}$ & 0 &  $-1$ & $0$ & $1-r$
\\
$V_{\pm}$ & 0 & 0 & 0 & 0 & $N_f$ & $\pm 1$ & $-(1-r)N_f + (N_c +1)$
\\
\hline
$v_{\pm}$ & 0 & 0 & 0 & 0 & $-N_f$ & $\pm 1$ & $(1-r)N_f - (N_c -1)$
\\
$M_{i}^{j}$ & 0 & 0 & $[0,\dots,0,1]_u$ & $[1,0,\dots,0]_{\tilde{u}}$ & $2$ & $0$ & $2r$
\\
\hline
\end{tabular}
}
\caption{
The gauge and global charges of the $U(N_f - N_c)$ theory with $N_f$ flavors (theory B).
\label{tfields2}
}
\end{table}

Theory B has a superpotential of the form
\beal{es0a55}
W= \widetilde{q}^{i}_{a} M_{i}^{j} q_{j}^{a} + v_{+} V_{-} + v_{-} V_{+} ~.~
\eea
The superpotential above gives the following F-term relations,
\beal{es0a56}
q \widetilde{q} = 0 ~,~ V_{+} = 0 ~,~ V_{-} = 0 ~,~
\eea
which imply that the singlets $V_{\pm}$ do not contribute to the moduli space of theory B. Furthermore, there are F-terms
\beal{es0a57}
\widetilde{q} M = 0 ~,~
M q = 0 ~,~
\eea
Overall, the relevant gauge invariant quantities that contribute to the moduli space of the dual theory are the singlets $M_{i}^{j}$ and the monopoles $v_{\pm}$. These are precisely the gauge invariant mesonic operators and bare monopole operators of theory A and follow from the identifications made by Aharony duality.
\\

\paragraph{IR free theories.} Let us comment on the case when $N_c = N_f$. The original $U(N_c)$ theory with $N_f = N_c$ flavors has a dual description, the theory of ${N_f}^2+2$ chiral multiplets with the superpotential
\beal{eq:v+v-M}
W = -v_+ v_- \det M ~.~
\eea
When $N_f = N_c = 1$, $U(1)$ theory A and its dual $XYZ$ theory B flow to the same interacting IR fixed point. On the other hand, when $N_c = N_f > 1$, it is known that the $U(N_c)$ theory A and its dual theory B flow to a IR free theory.

Firstly, let's consider the $N_c = N_f = 2$ case. As shown in \tref{tfields2}, the $R$-charges of $v_\pm$ and $M_i^j$ are $1-r N_f$ and $2 r$ respectively. $r$ is a parameter to be determined so that $1-r N_f$ and $2 r$ give correct $R$-charges at the IR fixed point. These $R$-charges are constrained by unitarity of the SCFT to be larger than 1/2 for interacting fields or to be equal to 1/2 for non-interacting fields. For $N_c = N_f = 2$, in order to meet the unitarity constraint one has $1-2 r =1/2$ and $2 r = 1/2$, which in turn indicates that $v_\pm$ and $M_i^j$ are non-interacting. Therefore, the $U(2)$ theory with two flavors, and its dual B theory, flow to a free theory in the IR. 

For $N_c = N_f > 2$, the situation is more complicated because both $R$-charges $1-r N_f$ and $2 r$ cannot be larger than or equal to 1/2 simultaneously.\footnote{In general, this happens for cases which do not satisfy $N_f > \frac{4N_c -2}{3}$. Such theories have been studied in \cite{Safdi:2012re}. In this work, we only study cases for which the unitary bound is not broken.} This however doesn't mean that the unitarity bound cannot be met. Instead, new $U(1)$ symmetry emerge in IR and the $R$-charges would get corrections from the new symmetry to meet the unitarity constraint for a IR fixed point.  One can understand this better with theory B and the reader is referred to \cite{Willett:2011gp,Bashkirov:2011vy,Benini:2011mf,Hwang:2012jh}. 
\\

\paragraph{Towards the algebraic structure of the Moduli Spaces.}
The following sections focus on theory A and refer to theory B via Aharony duality. The focus is to identify the algebraic structure of the moduli spaces by computing the \textit{Hilbert series} \cite{Benvenuti:2006qr,Hanany:2006uc,Feng:2007ur,Butti:2007jv,Hanany:2007zz,Forcella:2008bb,Forcella:2008eh} for theory A. The Hilbert series counts gauge invariant operators that characterizes the entire chiral ring. By direct generalization from the 3d $\mathcal{N}=4$ theories, the monopole operators for 3d $\mathcal{N}=2$ theories are dressed by gauge invariant operators which are invariant under the residual gauge symmetry left unbroken in the monopole background. The following section outlines the computation of the Hilbert series which counts dressed monopole operators for 3d $\mathcal{N}=2$ theories.
\\

%==================================================%
%==================================================%
\section{Hilbert Series \label{s2}}

%==================================================%
\subsection{Computation \label{sgno}}

The Hilbert series counts gauge invariant operators on the moduli space of a supersymmetric gauge theory. By doing so, the Hilbert series identifies the algebraic structure of the moduli space of the theory. For the $3d$ $\mathcal{N}=2$ theory with $U(N_c)$ gauge group and $N_f$ flavors, the Hilbert series counts mesonic gauge invariant operators of the form $M=Q\tilde{Q}$ on the Higgs branch and dressed monopole operators of the form $M_m v_m$ on the remaining moduli space of the theory. The aim of this section is to introduce the computation of the Hilbert series for the $3d$ $\mathcal{N}=2$ theory.
\\

\paragraph{Conformal dimension of monopole operators.} For the $3d$ $\mathcal{N}=2$ theory with gauge group $U(N_c)$ and $N_f$ flavors, the conformal dimension of a monopole operator with GNO charge $m=(m_1,\dots, m_{N_c})$ has the general form
\beal{es0a20}
\Delta(m) = N_f (1-r) \sum_{i}^{N_c} |m_i| - \sum_{i<j} |m_i - m_j| ~,~
\eea
where $r$ is the $U(1)_R$ charge of $Q_{i}^{a}$. As reviewed in section \sref{s1}, instanton effects \cite{Aharony:1997gp,Aharony:1997bx,Karch:1997ux,Aharony:2013dha} lift most of the moduli space such that the remaining monopole operators carry only magnetic charges $m_1 \geq 0 \geq m_2$. Accordingly, \eref{es0a20} simplifies for $N_c>1$ to
\beal{es0a21}
\Delta(m_1,m_2) &=& 
((1- r) N_f-(N_c-1)) (m_1 - m_2)
~.~
\eea
If $N_c=1$, the conformal dimension is
\beal{es0a22}
\Delta(m) = (1-r) N_f |m| ~,~
\eea
where $m\in \mathbb{Z}$.
\\

\paragraph{Hilbert Series formula.} The Hilbert series for the $U(N_c)$ theory with $N_f$ flavors is given by \cite{Cremonesi:2013lqa}
\beal{es0b1}
&&
g(\textbf{t},\tau,a,u,\tilde{u}; \mathcal{M}_{U(N_c), N_f}) =
\nn\\
&&
\hspace{1cm}
\sum_{m_1 = 0}^\infty \sum_{m_2=-\infty}^0 \tau^{J(m_1,m_2)} a^{K(m_1,m_2)} \textbf{t}^{\Delta(m_1,m_2)} 
P_{U(N_c)}(m_1,m_2; {}u,\tilde{{}u},\textbf{t}) 
~,~
\nn\\
\eea
where $\textbf{t}$ counts the monopole operators according to their conformal dimension. $J(m_1,m_2)=m_1+m_2$ and $K(m_1,m_2)= - N_f (m_1-m_2)$ are respectively the charges under the $U(1)_T$ topological and $U(1)_A$ axial symmetries. The respective fugacities are chosen to be $\tau$ and $a$. The above Hilbert series is further refined under the flavour symmetries $SU(N_f)_1$ and $SU(N_f)_2$ with the fugacities ${}u$ and $\tilde{{}u}$ respectively.

Instead of using fugacity $\textbf{t}$, one can identify a fugacity basis in terms of a new $U(1)$ symmetry that weights the bare monopole operators $v_{+}$ and $v_{-}$ and mesonic operators $M_m$ equally. By doing so, a new fugacity $t$ corresponding to this new $U(1)$ symmetry can be introduced which counts degrees of chiral operators according to the number of $v_+$, $v_-$ and $M_m$. The fugacity map between $\textbf{t}$ and $t$ is as follows,
\beal{es0b30}
t = \textbf{t}^{\frac{2(N_f-N_c+1)}{N_f+2}} ~,~
\eea
with $r$ mapping to the value $r\mapsto r_{0}=\frac{(N_f-N_c+1)}{N_f+2}$ under the new $U(1)$ symmetry.
In the following sections, fugacity $t$ is used instead of $\textbf{t}$ in the Hilbert series.

The dressing of monopole operators comes from the classical factor $P_{U(N_c)}(m_1,m_2; {}u,\tilde{{}u}, t) $ in \eref{es0b1}. As discussed in section \sref{s1}, depending on the magnetic charge of the monopole operator, the gauge group $U(N_c)$ is broken to a residual subgroup $H_m \subset U(N_c)$. The dressing factor is a separate Hilbert series which counts mesonic operators of the form $M_m = Q \tilde{Q}$ which are invariant under the residual subgroup $H_m$. It takes the form \cite{Gray:2008yu}
\beal{es0a26}
P_{U(N_c)}(m_1,m_2;{}u,\tilde{{}u},t) &=& \oint \ud\mu_{H_m \subseteq U(N_c)}
~ \PE\Big[
[1,0,\dots,0]_z [0,\dots,0,1]_{{}u} w a t^{1/2} +
\nn\\
&&
\hspace{1cm}
[0,\dots,0,1]_z [1,0,\dots,0]_{\tilde{{}u}} w^{-1} a t^{1/2}
\Big] ~,~
\eea
where $\ud\mu_{H_m}$ is the Haar measure of $H_m$ and fugacities $z$ and $w$ correspond respectively to the non-Abelian subgroup of $H_m$ and a $U(1)$ factor of $H_m$. The remaining $U(1)$ factors in $H_m$ do not give charge to the matter fields. The dressing factor takes a concise form when one uses the highest weight generating function of characters of the flavour symmetry $SU(N_f)_1 \times SU(N_f)_2$. It is
\beal{es0b100}
\mathcal{F}_{N_c,N_f} = 
\PE\left[
\sum_{i=1}^{N_c}
\mu_{N_f - i} \nu_{i} a^{2i} t^{i r}
\right]
~,~
\eea
where $\mu_i, \nu_i$ count highest weights of $SU(N_f)$ representations. Monomials in $\mu_i, \nu_i$ are replaced by characters of $SU(N_f)$ 
\beal{es0b101}
\prod_{i=1}^{N_f-1} \mu_{i}^{n_i}  \mapsto [n_1, \dots , n_{N_f - 1}]_{{}u}^{SU(N_f)_1}
~~,~~
 \prod_{i=1}^{N_f-1} \nu_{i}^{n_i} \mapsto [n_1, \dots , n_{N_f - 1}]_{\tilde{{}u}}^{SU(N_f)_2} 
~.~
\eea
\\

\paragraph{Plethystic Logarithm.} The plethystic logarithm \cite{Benvenuti:2006qr,Feng:2007ur,Hanany:2007zz} of the Hilbert series $g(t;\mathcal{M}_{U(N_c),N_f})$ is defined as
\beal{es0a50}
\PL\left[
g(t;\mathcal{M}_{U(N_c),N_f})
\right]
= \sum_{k=1}^{\infty} \frac{\mu(k)}{k} \log\left[
g(t^k;\mathcal{M}_{U(N_c),N_f})
\right]
~,~
\eea
where $\mu(k)$ is the M\"obius function. The plethystic logarithm has a series expansion in $t$. It extracts information from the Hilbert series about the algebraic structure of the moduli space. As an expansion in $t$, the initial positive terms refer to generators of the moduli space. The following negative terms refer to first order relations amongst the generators. When the series terminates at this point, the moduli space is known to be a \textit{complete intersection} moduli space. If the series does not terminate, the moduli space is known to be a \textit{non-complete intersection} where relations form higher order relations known as \textit{syzygies} \cite{Benvenuti:2006qr,Feng:2007ur,Hanany:2007zz}. We expect the moduli space of the $U(N_c)$ theory with $N_f$ flavors to be in one of these two classes.
\\

%==================================================%
\subsection{Examples: $U(1)$ \label{sex1}}

%-------%
\subsubsection{$U(1)$ with 1 flavor: 3 identical components}

The Hilbert series is given by
\beal{es1a2}
g(\textbf{t};\mathcal{M}_{U(1),1}) = \sum_{m=-\infty}^{\infty} \textbf{t}^{\Delta(m)} P_{U(1)}(m;\textbf{t}) ~,~
\eea
where fugacity $\textbf{t}$ counts the bare monopole operators according to their conformal dimension.
For a $U(1)$ theory with $N_f=1$ the conformal dimension of the bare monopole operator is given by
\beal{es1a1}
\Delta(m) = (1-r) |m| ~,~
\eea
where $r$ is the $U(1)$ R-charge of the fundamental $Q$ and anti-fundamental $\widetilde{Q}$ and $m\in \mathbb{Z}$ is the GNO magnetic flux. Under a refinement, fugacities $\tau$ and $a$, which respectively correspond to the topological symmetry $U(1)_T$ and axial symmetry $U(1)_A$, can be added to the formula in \eref{es1a2}. The refined Hilbert series takes the form
\beal{es1a1b}
g(\textbf{t},\tau,a;\mathcal{M}_{U(1),1}) = \sum_{m=-\infty}^{\infty} \tau^{J(m)} a^{K(m)} \textbf{t}^{\Delta(m)} P_{U(1)}(m;\textbf{t},y) ~,~
\eea
where $J(m)=m$ and $K(m)=-|m|$ are respectively the topological and axial $U(1)$ charges of a monopole operator of GNO charge $m$.

As discussed in \eref{es0b30}, a new $U(1)$ symmetry can be introduced under which the monopole operators and mesonic operators are weighted equally. Under this new $U(1)$ symmetry, $r \mapsto r_0=\frac{1}{3}$ and the following fugacity map applies 
\beal{es1a6b}
t=\textbf{t}^{\frac{2}{3}} ~,~
\eea
where now $t$ counts the number of $v_{+}, v_{-}$ and $M_m=Q \tilde{Q}$.

In the Hilbert series formulas, the classical factor in its refined form is given by
\beal{es1a4}
P_{U(1)}(m; t,a)
=
\left\{
\ba{cc}
\oint_{|w|=1} \frac{\ud w}{w} ~\PE\left[
w a t^{1/2} + w^{-1} a t^{1/2}
\right]
 = \frac{1}{1-a^2 t} & \hspace{1cm} m = 0
\\
1 & \hspace{1cm} m \neq 0
\ea
\right.
~,~
\eea
where $w$ is the $U(1)$ gauge charge fugacity and $a$ the $U(1)_A$ axial charge fugacity.\footnote{The $U(1)$ R-charge was found to be $r=\frac{1}{3}$ in \cite{Aharony:1997bx}.}

\begin{figure}[ht!!]
\begin{center}
\includegraphics[trim=0cm 1cm 0cm 0cm,width=0.5\textwidth]{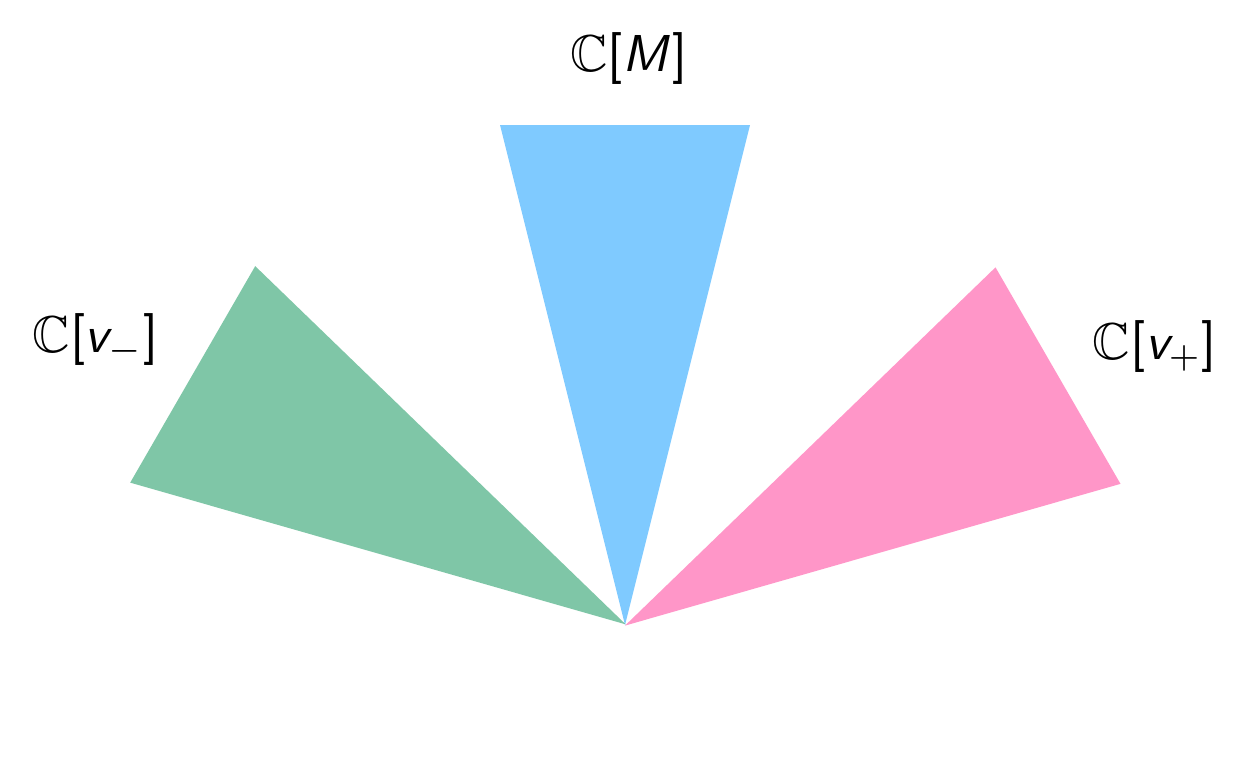}
\caption{
The moduli space of the $U(1)$ theory with $N_f=1$ is made of 3 1-dimensional cones $\mathbb{C}$ which meet at the origin. The cones are indistinguishable and are each generated by the mesonic generator and the monopole operators $v_{+}$ and $v_{-}$.
\label{fcone11}
}
 \end{center}
 \end{figure}

Summing up the refined series in \eref{es1a1b} gives
\beal{es1a6}
&&
g(t,\tau,a;\mathcal{M}_{U(1),1}) =\frac{1}{1- a^2 t} + \frac{1}{1-\tau a^{-1} t} +\frac{1}{1-\tau^{-1} a^{-1} t} - 2
\nn\\
&&
\hspace{1cm}
=
1 
+ a^2 t  + (\tau + \tau^{-1}) a^{-1} t 
+ a^4 t^2 + (\tau^2 + \tau^{-2}) a^{-2} t^2
+ \dots
~.~
\eea

The plethystic logarithm of the Hilbert series is
\beal{es1a7}
\PL\left[
g(t,\tau,a;\mathcal{M}_{U(1),1})
\right]
&=&
a^2 t + (\tau + \tau^{-1}) a^{-1} t 
 - (\tau + \tau^{-1}) a t^{2}  - a^{-2} t^{2} + \dots ~.~
\nn\\
\eea
The generators corresponding to the first positive terms can be identified from the above plethystic logarithm as follows,
\beal{es1a8}
\ba{ccc}
\text{$\PL$ term} & \rightarrow & \text{generator}
\\
a^2 t & \rightarrow & M = Q^{a} \tilde{Q}_{a}
\\
 \tau a^{-1} t & \rightarrow & v_{+}
\\
 \tau^{-1} a^{-1} t  & \rightarrow & v_{-}
\ea
~~,~~
\eea
the meson $M$ and two monopoles $v_\pm$. The first order relations formed by the generators are as follows,
\beal{es1a9}
\ba{ccc}
\text{$\PL$ term} & \rightarrow & \text{relation}
\\
 - \tau a t^{2} & \rightarrow & v_{+} M = 0
\\
 - \tau^{-1} a t^{2}& \rightarrow & v_{-} M = 0
\\
- a^{-2} t^{2}  & \rightarrow & v_{+} v_{-} = 0
\ea
~~.~~
\eea

The full moduli space of the $U(1)$ theory with $N_f=1$ is an algebraic variety,
\beal{es1d10}
\mathcal{M}_{U(1),1} = 
\mathbb{C}[M,v_{+},v_{-}] / \langle
v_{+} M = 0 ~,~
v_{-} M = 0 ~,~
v_{+} v_{-} = 0 
\rangle~.~
\eea
The moduli space $\mathcal{M}_{U(1),1}$ has the following components
\beal{es1d11}
&
\mathcal{M}_{U(1),1}^{0} = \mathbb{C}[M]
~,~&
\nn\\
&
\mathcal{M}_{U(1),1}^{+} = \mathbb{C}[v_{+}] 
~,~
\mathcal{M}_{U(1),1}^{-} = \mathbb{C}[v_{-}] 
~,~
&
\eea
where $\mathcal{M}_{U(1),1}^{0}$ is the Higgs branch and $\mathcal{M}_{U(1),1}^{+}$ and $\mathcal{M}_{U(1),1}^{-}$ are the Coulomb branches of the theory. 
The corresponding Hilbert series are 
\beal{es1d12}
&g(t,\tau,a;\mathcal{M}_{U(1),1}^{0}) = \frac{1}{1-a^2 t} ~,~&
\nn\\
&
g(t,\tau,a;\mathcal{M}_{U(1),1}^{+}) = \frac{1}{1-\tau a^{-1} t} ~,~
g(t,\tau,a;\mathcal{M}_{U(1),1}^{-}) = \frac{1}{1-\tau^{-1} a^{-1} t} ~.~
&
\eea
The moduli space $\mathcal{M}_{U(1),1}$ is made of 3 identical cones $\mathbb{C}$ which meet at the origin, as shown in \fref{fcone11}. 

The moduli space is the union of the 3 components. By removing contributions from the intersections, the Hilbert series of the full moduli space can therefore be expressed as
\beal{es1d13}
g(t,\tau,a;\mathcal{M}_{U(1),1})  = g(t,\tau,a;\mathcal{M}_{U(1),1}^{0}) + g(t,\tau,a;\mathcal{M}_{U(1),1}^{+}) + g(t,\tau,a;\mathcal{M}_{U(1),1}^{-})  - 2 
~,~
\nn\\
\eea
where the intersections at the origin are taken care of by $-2$.
\\

%-------%
\subsubsection{$U(1)$ with 2 flavors: 3 components with non-Abelian symmetry}

The Hilbert series for the $U(1)$ theory with 2 flavors is given by
\beal{es2a2}
g(\textbf{t};\mathcal{M}_{U(1),2}) = \sum_{m=-\infty}^{\infty} \textbf{t}^{\Delta(m)} P_{U(1)}(m;\mathbf{t}) ~,~
\eea
where $\textbf{t}$ is the fugacity which counts bare monopole operators according to their conformal dimension.
For a $U(1)$ theory with $N_f=2$ the conformal dimension of the bare monopole operator is given by
\beal{es2a1}
\Delta(m) = 2(1-r) |m| ~,~
\eea
where $r$ is the $U(1)$ R-charge of the fundamental $Q_i$ and anti-fundamental $\widetilde{Q}^i$ and $m\in \mathbb{Z}$ is the GNO magnetic flux. The Hilbert series formula above can be refined with the charges from the topological symmetry $U(1)_T$ and the axial symmetry $U(1)_A$. The respective fugacities are chosen to be $\tau$ and $a$. The refined Hilbert series is 
\beal{es2a2b}
g(\textbf{t},\tau,a,u,\tilde{u};\mathcal{M}_{U(1),2}) = \sum_{m=-\infty}^{\infty} \tau^{J(m)} a^{K(m)} \textbf{t}^{\Delta(m)} P_{U(1)}(m;\textbf{t},a,u,\tilde{u}) ~,~
\eea
where $J(m)=m$ and $K(m)=-2|m|$ are respectively the topological and axial charges of a monopole operator with GNO charge $m$ as discussed in \tref{tfields}. Under a new $U(1)$ symmetry that weights monopole operators $v_{\pm}$ and mesonic operators $M_m$ equally, a new fugacity $t$ can be introduced by mapping the value of $r$ to $r\mapsto r_0 = \frac{1}{2}$. As discussed in \eref{es0b30}, the fugacity map is $ t=\textbf{t}$.

The classical factor of the Hilbert series formula is $P_{U(1)}(m;t,a,u,\tilde{u})$ and it is further refined under the flavour symmetries $SU(2)_1\times SU(2)_2$. The fugacities $u$ and $\tilde{u}$ respectively count charges under $SU(2)_1$ and $SU(2)_2$. The refined classical factor is given by
\beal{es2a4}
&&
P_{U(1)}(m;t,a,u,\tilde{u})
=
\nn\\
&&
\hspace{1cm}
\left\{
\ba{cc}
\oint_{|w|=1} \frac{\ud w}{w}~ \PE\left[
w (u+u^{-1})  a t^{1/2} + w^{-1} (\tilde{u} + \tilde{u}^{-1}) a t^{1/2}
\right]
= f
& \hspace{1cm} m = 0
\\
1 & \hspace{1cm} m \neq 0
\ea
\right.
~,~
\nn\\
\eea
where the integral gives
\beal{es2a5b}
f= \frac{
(1- a^4 t^2)
}{
(1- u \tilde{u} a^2 t)
(1- u \tilde{u}^{-1} a^2 t)
(1- u^{-1} \tilde{u} a^2 t)
(1- u^{-1} \tilde{u}^{-1} a^2 t)
}
~.~
\eea
From the above Hilbert series corresponding to the classical component of the moduli space where the GNO magnetic flux is $m=0$, one can identify the classical component to be the conifold $\mathcal{C}$. The 4 generators of the conifold are the mesonic operators $M=Q \tilde{Q}$ which satisfy the quadratic relation $\det{M} =0 $.

\begin{figure}[ht!!]
\begin{center}
\includegraphics[trim=0cm 1cm 0cm 0cm,width=0.5\textwidth]{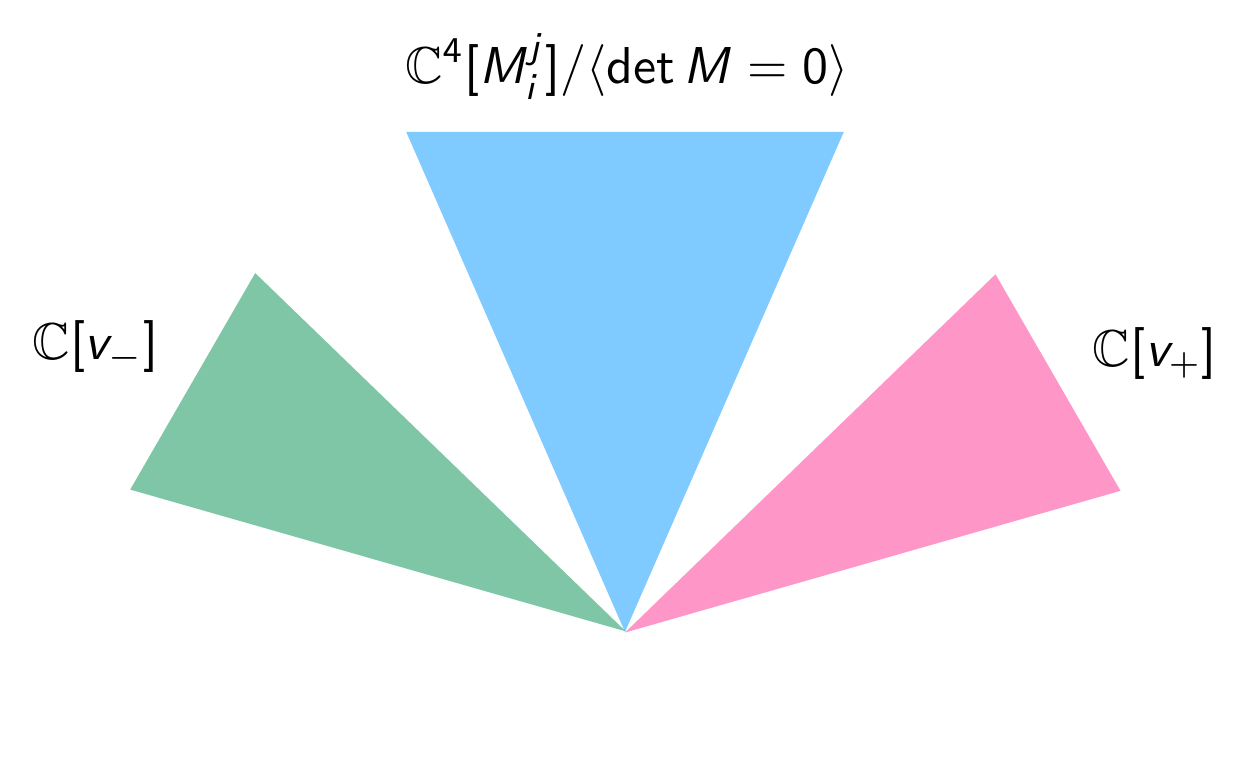}
\caption{
The moduli space of the $U(1)$ theory with $N_f=2$ is made of 3 cones which meet at the origin. 
\label{fcone12}
}
 \end{center}
 \end{figure}
 
Summing up the Hilbert series formula in \eref{es2a2} for the entire moduli space gives
\beal{es2a6}
g(t,\tau,a,u,\tilde{u};\mathcal{M}_{U(1),2}) &=&
f
+
\frac{1}{1-\tau a^{-2} t} + \frac{1}{1-\tau^{-1} a^{-2} t} - 2
~.~
\eea
From the Hilbert series above one can observe that the moduli space is made of 3 cones, one being the conifold generated by the mesonic operators and the other two being two $\mathbb{C}$, each generated by monopole operators of opposite topological $U(1)_T$ charge. The 3 cones meet at the origin as shown in \fref{fcone12}.
\\

The Hilbert series has the following character expansion,
\beal{es2a7b}
&&
g(t,\tau,a,u,\tilde{u};\mathcal{M}_{U(1),2})
= \sum_{n=0}^{\infty}
\Big[
[n]_{u} [n]_{\tilde{u}} a^{2n}
+ (\tau^{n}+\tau^{-n}) a^{-2n}
\Big]
t^{n}
-2
\nn\\
&&
\hspace{1cm}
=
1
+ [1]_u [1]_{\tilde{u}} a^2 t
+ (\tau + \tau^{-1}) a^{-2} t
+ [2]_u [2]_{\tilde{u}} a^4 t^2
+ (\tau^2 + \tau^{-2}) a^{-4} t^2
+ \dots
~.~
\nn\\
\eea
The plethystic logarithm of the refined Hilbert series of the full moduli space is
\beal{es2a7}
&&
\PL\left[
g(t,\tau,a,u,\tilde{u};\mathcal{M}_{U(1),2})
\right]
=
[1]_{u} [1]_{\tilde{u}} a^2 t
+ (\tau+ \tau^{-1}) a^{-2} t
\nn\\
&&
\hspace{5cm}
- a^4 t^{2} 
- a^{-4} t^{2}
- [1]_{u} [1]_{\tilde{u}} (\tau + \tau^{-1}) t^{2}
+ \dots ~.~
\eea
From the initial positive terms of the plethystic logarithm, one can identify the generators of the moduli space,
\beal{es2a8}
\ba{ccc}
\text{$\PL$ term} & \rightarrow & \text{generator}
\\
+ [1]_{u} [1]_{\tilde{u}} a^2 t  & \rightarrow & M_{i}^{j} = Q_{i}^{a} \tilde{Q}_{a}^{j}
\\
+ \tau a^{-2} t & \rightarrow & v_+
\\
+ \tau^{-1} a^{-2} t & \rightarrow & v_-
\ea
~~.~~
\eea
The generators are the mesons and the bare monopoles. The first order relations formed among the generators are identified as follows,
\beal{es2a9}
\ba{ccc}
\text{$\PL$ term} & \rightarrow & \text{relation}
\\
- a^4 t^2  & \rightarrow & \det{M} = 0
\\
- [1]_{u} [1]_{\tilde{u}} \tau t^{2} & \rightarrow & v_{+} M_{i}^{j} = 0
\\
- [1]_{u} [1]_{\tilde{u}} \tau^{-1} t^{2} & \rightarrow & v_{-} M_{i}^{j} = 0
\\
- a^{-4} t^2 & \rightarrow & v_+ v_- = 0
\ea
~~.~~
\eea

The full moduli space of the $U(1)$ theory with $N_f=2$ can be expressed as the following algebraic variety,
\beal{es2a9b}
\mathcal{M}_{U(1),2} 
=
\mathbb{C}[M_i^j, v_{+}, v_{-}] / \langle
\det M = 0 ~,~
v_{+} M_i^j = 0 ~,~
v_{-} M_i^j = 0 ~,~
v_+ v_- = 0 
\rangle
~.~
\eea
The moduli space $\mathcal{M}_{U(1),2}$ has the following components
\beal{es2d1}
&
\mathcal{M}_{U(1),2}^{0} = \mathbb{C}[M_i^j] / \langle \det{M} = 0 \rangle = \mathcal{C}~,~
&
\nn\\
&
\mathcal{M}_{U(1),2}^{+} = \mathbb{C}[v_+] ~,~
\mathcal{M}_{U(1),2}^{-} = \mathbb{C}[v_-] ~,~
&
\eea
where the Higgs branch is given by $\mathcal{M}_{U(1),2}^{0}$ and the Coulomb branch by $\mathcal{M}_{U(1),2}^{+}$ and $\mathcal{M}_{U(1),2}^{-}$. The Higgs branch $\mathcal{M}_{U(1),2}^{0}$ is the conifold $\mathcal{C}$. The Hilbert series of the 3 components are as follows,
\beal{es2d2}
&
g(t,\tau,a,u,\tilde{u}; \mathcal{M}_{U(1),2}^{0} ) = \PE\left[ [1]_u [1]_{\tilde{u}} a^2 t  -a^4 t^2\right] ~,~
&
\nn\\
&
g(t,\tau,a,u,\tilde{u}; \mathcal{M}_{U(1),2}^{+} ) = \frac{1}{1-\tau a^{-2} t}~,~
g(t,\tau,a,u,\tilde{u}; \mathcal{M}_{U(1),2}^{-} ) = \frac{1}{1-\tau^{-1} a^{-2} t}~.~
&
\eea
The 3 components of the moduli space intersect only at the origin. 

The moduli space is the union of the 3 components. 
By removing the contributions from the intersections, the Hilbert series of  $\mathcal{M}_{U(1),2}$ therefore can be expressed as 
\beal{es2d3}
g(t,\tau,a,u,\tilde{u}; \mathcal{M}_{U(1),2} ) 
&=&
g(t,\tau,a,u,\tilde{u}; \mathcal{M}_{U(1),2}^{0} )  +
g(t,\tau,a,u,\tilde{u}; \mathcal{M}_{U(1),2}^{+} )  +
\nn\\
&&
g(t,\tau,a,u,\tilde{u}; \mathcal{M}_{U(1),2}^{-} )  - 2
~.~
\eea
\\

%==================================================%
%==================================================%
\subsection{Examples: $U(2)$ and $U(3)$ \label{sex2} \label{sex3}}

\subsubsection{$U(2)$ with 3 flavors: 4 components (Higgs, Mixed and Coulomb)}

The Hilbert series for the $U(2)$ theory with 3 flavors is given by
\beal{es21a1}
g(\textbf{t};\mathcal{M}_{U(2),3}) = \sum_{m_1 = 0}^\infty \sum_{m_2=-\infty}^0 \textbf{t}^{\Delta(m_1,m_2)} P_{U(2)}(m_1,m_2;\textbf{t}) ~,~
\eea
where fugacity $\textbf{t}$ counts bare monopole operators according to their conformal dimension. For the $U(2)$ theory with $N_f=3$ the conformal dimension of the bare monopole operator is given by
\beal{es21a3}
\Delta(m_1,m_2) = 
(2-3r) (m_1 - m_2)
%3(1-r) (m_1- m_2) - |m_1-m_2| 
~,~
\eea
where $m_1,m_2$ are the GNO magnetic fluxes. 

The Hilbert series expression in \eref{es21a1} can be refined to include fugacities $\tau$ and $a$ which respectively count charges of the topological $U(1)_T$ and axial $U(1)_A$ symmetries. The refined Hilbert series takes the following form
\beal{es21a3b}
&&
g(\textbf{t},\tau,a,u,\tilde{u};\mathcal{M}_{U(2),3}) = 
\nn\\
&&
\hspace{1cm}
\sum_{m_1 = 0}^\infty \sum_{m_2=-\infty}^0 
\tau^{J(m_1,m_2)} a^{K(m_1,m_2)}
\textbf{t}^{\Delta(m_1,m_2)} P_{U(2)}(m_1,m_2;\textbf{t},a,u,\tilde{u}) ~,~
\eea
where $J(m_1,m_2)= m_1+m_2$ and $K(m_1,m_2)= -3 (m_1- m_2)$ are respectively the topological and axial charges of a monopole operator with GNO charge $m_1,m_2$. In addition, the Hilbert series above is refined under the flavour symmetry $SU(3)_1\times SU(3)_2$, where fugacities $u$ and $\tilde{u}$ count the charges of the respective symmetries as summarised in \tref{tfields}. By introducing a new $U(1)$ symmetry that replaces $U(1)_R$ and weights monopole operators $v_{\pm}$ and mesonic operators $M_{m}$ equally, a new fugacity $t$ can be introduced that replaces $\textbf{t}$ by mapping the value of $r$ to $r\mapsto r_0 = \frac{2}{5}$. Following \eref{es0b30}, the fugacity map is $t= \textbf{t}^{\frac{4}{5}}$.

The classical contribution comes from the factor $P_{U(2)}(m_1,m_2; t,a,u,\tilde{u})$. The GNO charge lattice with $m_1,m_2$ can be dividend into 4 sublattices under which monopole operators that contribute to the moduli space are charged. Depending on which GNO sublattice one is, the gauge symmetry is either broken or unbroken. Accordingly, the classical factor of the Hilbert series can be written as follows,
\beal{es21a4}
&&
P_{U(2)}(m_1,m_2;t,a,u,\tilde{u})
\nn\\
&&
=
\left\{
\ba{ccc}
\oint \ud\mu_{SU(2)} \oint \ud\mu_{U(1)} \PE\left[
[1]_z w  [0,1]_{u} a t^{1/2} + [1]_z w^{-1} [1,0]_{\tilde{u}} a t^{1/2}
\right] = f_1 
& \hspace{0cm} m_1,m_2=0
\\
\oint \ud\mu_{U(1)}
\PE\left[
w [0,1]_{u} a t^{1/2} + w^{-1}  [1,0]_{\tilde{u}} a t^{1/2}
\right]
= f_2
& \hspace{0cm} \left\{\ba{c} m_1\neq 0, m_2 = 0 \\ m_1 = 0, m_2 \neq 0 \ea \right.
\\
1
& \hspace{0cm} m_1,m_2\neq 0
\ea
\right.
~,~
\nn\\
\eea
where the integrals above give
\beal{es21a4b}
f_1 &=&
\PE\left[
[0,1]_{u} [1,0]_{\tilde{u}} a^2 t - a^6 t^3
\right]
~,~
\nn\\
f_2 &=&
 (
 1 - [0,1]_{u} [1,0]_{\tilde{u}} a^4 t^2 + [1,1]_{u} a^6 t^3 + [1,1]_{\tilde{u}} a^6 t^3 - [0,1]_{u} [1,0]_{\tilde{u}} a^8 t^4
 + a^{12} t^6
 )
 \nn\\
 &&
 \times  
 \PE\left[ [0,1]_{u} [1,0]_{\tilde{u}} a^2 t \right] 
 ~.~
\eea

Summing up the refined Hilbert series in \eref{es21a3b} gives
\beal{es21a10}
&&
g(t,\tau,a,u,\tilde{u};\mathcal{M}_{U(2),3}) =
f_1 + f_2 \left[ \frac{1}{1-\tau a^{-3} t } + \frac{1}{1-\tau^{-1} a^{-3} t}  - 2\right] 
\nn\\
&&
\hspace{7cm}
+  
\frac{ a^{-6} t^2}{(1- \tau a^{-3} t) (1- \tau^{-1} a^{-3} t)}
~.~
\eea
The first few orders of the expansion of the Hilbert series is as follows,
\beal{es21a10c}
&&
g(t,\tau,a,u,\tilde{u};\mathcal{M}_{U(2),3}) =
1+ 
(
[0,1]_{u} [1,0]_{\tilde{u}} a^2 
+  (\tau + \tau^{-1}) a^{-3}
) t
\nn\\
&&
\hspace{0.2cm}
+ 
\big(
([0,2]_{u} [2,0]_{\tilde{u}} 
+ [0,1]_{u} [1,0]_{\tilde{u}} )a^4
+ [0,1]_{u} [1,0]_{\tilde{u}} (\tau + \tau^{-1}) a^{-1}
+ (\tau^{2} + \tau^{-2} + 1)  a^{-6}
\big) t^2
\nn\\
&&
\hspace{0.2cm}
+
\big(
(
[0,3]_{u} [3,0]_{\tilde{u}} 
+ [1,1]_{u} [1,1]_{\tilde{u}}
) a^6
+ [0,2]_{u} [2,0]_{\tilde{u}} (\tau + \tau^{-1}) a
\nn\\
&&
\hspace{0.7cm}
+ [1,1]_{u} [1,1]_{\tilde{u}} (\tau^2 + \tau^{-2}) a^{-4}
+ (\tau^3 + \tau + \tau^{-1} + \tau^{-3}) a^{-9}
\big)
t^3
+ \dots
~.~
\eea
The corresponding plethystic logarithm is
\beal{es21a12}
&&
\PL\left[
g(t,\tau,a,u,\tilde{u};\mathcal{M}_{U(2),3})
\right]
=
[0,1]_{u} [1,0]_{\tilde{u}} a^2 t
+ (\tau + \tau^{-1})a^{-3} t
\nn\\
&&
\hspace{2cm}
- a^6 t^{3}
-  [1,0]_{u} [0,1]_{\tilde{u}} (\tau + \tau^{-1} ) a t^{3} 
- [0,1]_{u} [1,0]_{\tilde{u}} a^{-4} t^{3} 
+ \dots ~.~
\eea

The plethystic logarithm encodes the generators and relations amongst generators which define the moduli space. The generators of the moduli space correspond to the initial positive terms of the plethystic logarithm. The generators are as follows
\beal{es21a13}
\ba{ccc}
\text{PL term} & \rightarrow & \text{generator}
\\
+ [0,1]_{u} [1,0]_{\tilde{u}} a^2 t  & \rightarrow & M_{i}^{j} = Q_{i}^{a} \tilde{Q}_{a}^{j}
\\
+ \tau a^{-3} t  & \rightarrow & v_+
\\
+ \tau^{-1} a^{-3} t    & \rightarrow & v_-
\ea
~~,~~
\eea
where $i,j=1,2,3$.
The corresponding first order relations between the generators are identified as follows
\beal{es21a15}
\ba{ccc}
\text{PL term} & \rightarrow & \text{relation}
\\
- a^6 t^{3} & \rightarrow &
\det{M} = 0
\\
-  [1,0]_{u} [0,1]_{\tilde{u}} \tau a t^{3}  & \rightarrow &
v_{+} {R_{(2,3)}}^{i}_{j} = 0
\\
-  [1,0]_{u} [0,1]_{\tilde{u}} \tau^{-1} a t^{3}  & \rightarrow &
v_{-} {R_{(2,3)}}^{i}_{j} = 0
\\
- [0,1]_{u} [1,0]_{\tilde{u}} a^{-4} t^{3} & \rightarrow &
v_+ v_- M_{i}^{j} = 0
\ea
~~,~~
\eea
where
\beal{es21a15b}
{R_{(2,3)}}^{i}_{j} = \frac{1}{2}  \epsilon_{j k_1 k_2} \epsilon^{i m_1 m_2} M_{m_1}^{k_1} M_{m_2}^{k_2}
~.~
\eea

From the generators and first order relations, the moduli space can be expressed as the following algebraic variety,
\beal{es21a16}
\mathcal{M}_{U(2),3} = \mathbb{C}[M_{i}^{j},v_{\pm}] /
\langle
\det{M} = 0 ,
%v_{\pm} \text{minors}(2,3) = 0 ,
v_{\pm} {R_{(2,3)}}^{i}_{j} = 0,
v_{+} v_{-} M_{i}^{j} = 0
\rangle
~.~
\eea
Let us call the space of $N\times N$ matrices $M_{i}^{j}$ with at most rank $k$ as $\mathcal{M}_{k,N}$. Using this space, the components of the moduli space $\mathcal{M}_{U(2),3}$ can be expressed as  
\beal{es21d1}
&
\mathcal{M}_{U(2),3}^{0} = 
\mathcal{M}_{2,3}
%\mathbb{C}[M_{i}^{j}] / \langle \det{M} = 0 \rangle ~,~
&
\nn\\
&
\mathcal{M}_{U(2),3}^{+} = 
\mathcal{M}_{1,3}  \times \mathbb{C}[v_{+}]~,~
%\mathbb{C}[M_{i}^{j}] / \langle {R_{(2,3)}}^{i}_{j} = 0 \rangle  \times \mathbb{C}[v_{+}]~,~
\mathcal{M}_{U(2),3}^{-} = 
\mathcal{M}_{1,3}  \times \mathbb{C}[v_{-}]~,~
%\mathcal{M}_{U(2),3}^{-} = \mathbb{C}[M_{i}^{j}] / \langle {R_{(2,3)}}^{i}_{j} = 0 \rangle  \times \mathbb{C}[v_{-}]~,~
&
\nn\\
&
\mathcal{M}_{U(2),3}^{+-} = \mathbb{C}[v_{+},v_{-}]~,~
&
\eea
where $\mathcal{M}_{U(2),3}^{0}$ and $\mathcal{M}_{U(2),3}^{+-} $ are identified as Higgs and Coulomb branches respectively while $\mathcal{M}_{U(2),3}^{+} $ and $\mathcal{M}_{U(2),3}^{-} $ are mixed branches. The corresponding Hilbert series are as follows,
\beal{es21d2}
&&
g(t,\tau,a,u,\tilde{u}; \mathcal{M}_{U(2),3}^{0}) 
=
f_1
~,~
\nn\\
&&
g(t,\tau,a,u,\tilde{u}; \mathcal{M}_{U(2),3}^{+}) 
=
f_2 \times \frac{1}{1 - \tau a^{-3} t}
 ~,~
 \nn\\
 &&
g(t,\tau,a,u,\tilde{u}; \mathcal{M}_{U(2),3}^{-}) 
=
f_2 \times \frac{1}{1 - \tau^{-1} a^{-3} t}
 ~,~
 \nn\\
 &&
g(t,\tau,a,u,\tilde{u}; \mathcal{M}_{U(2),3}^{+-}) 
=
\frac{1}{(1 - \tau a^{-3} t)(1 - \tau^{-1} a^{-3} t)}
 ~,~
 \eea
 where $f_1$ and $f_2$ correspond to the monopole dressing factors in \eref{es21a4b}.
 The 4 components intersect in various subspaces which are
 \beal{es21d3}
  \mathcal{I}_{0} = \{  0 \}
 ~,~
 \mathcal{I}_{M} = \mathcal{M}_{1,3}
 ~,~
 \mathcal{I}_{+} = \mathbb{C}[v_{+}]
 ~,~
 \mathcal{I}_{-} = \mathbb{C}[v_{-}]
 ~,~
 \eea
 where $\{0\}$ is the origin. The corresponding Hilbert series are
 \beal{es21d4}
 &
 g(t,\tau,a,u,\tilde{u};\mathcal{I}_{M} ) = f_2 ~,~
 g(t,\tau,a,u,\tilde{u}; \mathcal{I}_{0} ) = 1 ~,~
&
\nn\\
&
g(t,\tau,a,u,\tilde{u}; \mathcal{I}_{+} ) = \frac{1}{1 - \tau a^{-3} t} ~,~
g(t,\tau,a,u,\tilde{u}; \mathcal{I}_{-} ) = \frac{1}{1 - \tau^{-} a^{-3} t} ~.~
&
 \eea
 
The union of the 4 components is the moduli space.  
By removing contributions from the intersections, the Hilbert series of the full moduli space $\mathcal{M}_{U(2),3}$ can be expressed as
\beal{es21d5}
&&
g(t,\tau,a,u,\tilde{u}; \mathcal{M}_{U(2),3}) 
=
g(t,\tau,a,u,\tilde{u}; \mathcal{M}_{U(2),3}^{0}) + 
g(t,\tau,a,u,\tilde{u}; \mathcal{M}_{U(2),3}^{+}) 
\nn\\
&&
\hspace{2cm}
+
g(t,\tau,a,u,\tilde{u}; \mathcal{M}_{U(2),3}^{-}) +
g(t,\tau,a,u,\tilde{u}; \mathcal{M}_{U(2),3}^{+-})  
- 2 g(t,\tau,a,u,\tilde{u}; \mathcal{I}_{M} )
\nn\\
&&
\hspace{2cm}
- g(t,\tau,a,u,\tilde{u}; \mathcal{I}_{+} ) 
- g(t,\tau,a,u,\tilde{u}; \mathcal{I}_{-} )
+  g(t,\tau,a,u,\tilde{u}; \mathcal{I}_{0} )
~.~
\eea
This expression for the Hilbert series of the full moduli space is in agreement with the Hilbert series expression in \eref{es21a10}.
\\

%---------------------%
\subsubsection{$U(3)$ with 4 flavors: 4 components (Higgs and Mixed)}

The Hilbert series for the $U(3)$ theory with $4$ flavors is given by
\beal{es23a1}
g(\textbf{t};\mathcal{M}_{U(3),4}) =
\sum_{m_1 = 0}^\infty \sum_{m_2=-\infty}^0 \textbf{t}^{\Delta(m_1,m_2,m_3 = 0)} P_{U(3)}(m_1,m_2; \textbf{t}) ~,~
\eea
where $\textbf{t}$ is the fugacity which counts bare monopole operators according to their conformal dimension. The following is the conformal dimension of the bare monopole operator for the $U(3)$ theory with $N_f=4$,
\begin{align}
\Delta(m_1,m_2) = (2-4r)(m_1 -m_2)
\end{align}
where $m_1,m_2$ are the GNO magnetic fluxes. Note that all monopole operators which contribute to the moduli spaces carry fluxes $m_1 \geq 0,m_2 \leq 0$ with $m_3 = 0$, as discussed in section \sref{sgno}. The Hilbert series in \eref{es23a1} can be refined to include fugacities which count charges under the topological $U(1)_T$ and axial $U(1)_A$ symmetry. The respective fugacities are chosen to be $\tau$ and $a$. Accordingly, the refined Hilbert series takes the form
\beal{es23a1b}
&&
g(\textbf{t},\tau,a,u,\tilde{u};\mathcal{M}_{U(3),4}) =
\nn\\
&&
\hspace{1cm}
\sum_{m_1 = 0}^\infty \sum_{m_2=-\infty}^0 \tau^{J(m_1,m_2} a^{K(m_1,m)2} \textbf{t}^{\Delta(m_1,m_2)} P_{U(3)}(m_1,m_2;\textbf{t},a,u,\tilde{u}) ~,~
\eea
where $J(m_1,m_2)=m_1+m_2$ and $K(m_1,m_2)=-4(m_1 - m_2)$ are respectively the topological and axial charges of a monopole operator with GNO charge $m_1,m_2$. The Hilbert series in \eref{es23a1b} is further refined by the flavour symmetry $SU(4)_1\times SU(4)_2$ whose charges are counted respectively by fugacities $u$ and $\tilde{u}$. A new $U(1)$ symmetry that replaces $U(1)_R$ can be introduced such that monopole operators $v_{\pm}$ and mesonic operators $M_m$ are counted equally by a new fugacity $t$. This new fugacity replaces $\textbf{t}$ by mapping the value of r to $r\mapsto r_0 = \frac{1}{3}$. The corresponding fugacity map is $t= \textbf{t}^{\frac{2}{3}}$.

The classical contribution to the Hilbert series comes from the factor $P_{U(3)}(m_1,m_2; t,a,u,\tilde{u}) $. The GNO charge lattice can be divided into 4 sublattices. Depending on which sublattice the GNO charge of a monopole operator is located, the gauge symmetry breaks under the Higgs mechanism. The residual gauge symmetry determines the dressing of the monopole operator in the particular GNO charge sublattice. Accordingly, the classical factor of the Hilbert series can be written as follows,
\beal{es23a5}
&&
\hspace{-0.5cm}
P_{U(3)}(m_1,m_2; t,a,u,\tilde{u}) 
=
\nn\\
&&
\hspace{-0.5cm}
\left\{
\ba{cc}
\oint \ud\mu_{SU(3)} \oint \ud\mu_{U(1)}
\PE\left[
  [1,0]_z w [0,0,1]_{u} a t^{1/2} + [0,1]_z  w^{-1} [1,0,0]_{\tilde{u}}  a t^{1/2}
\right]
= 
f_1
%\frac{1+p+p^2+p^3}{(1-p)^{15}}
& \hspace{0cm}
m_1,m_2 = 0
\\
\oint \ud\mu_{SU(2)} \oint \ud\mu_{U(1)}
\PE\left[
 [1]_z  w [0,0,1]_{u} a t^{1/2} + [1]_z  w^{-1} [1,0,0]_{\tilde{u}} a t^{1/2}
\right]
=
f_2
%\frac{1+4p+10p^2+4p^3+p^4}{(1-p)^{12}}
& \hspace{0cm}
\left\{ \ba{c} m_1\neq 0, m_2 = 0 \\ m_1 = 0, m_2 \neq 0 \ea \right.
\\
\oint \ud\mu_{U(1)}
\PE\left[
z  [0,0,1]_{u} a t^{1/2} + z^{-1} [1,0,0]_{\tilde{u}} a t^{1/2}
\right]
= 
f_3
%\frac{1+9p+9p^2+p^3}{(1-p)^7}
& \hspace{0cm}
m_1,m_2 \neq 0
\ea
\right.
~,~
\nn\\
\eea
where the integrals above give
\beal{es23a5b}
f_1
&=& 
\PE\left[
[0,0,1]_{u} [1,0,0]_{\tilde{u}} a^2 t- a^8 t^4
\right]
~,~
\nn\\
f_2
&=&
(
1
- [0,0,1]_{u} [1,0,0]_{\tilde{u}} a^6 t^3
+ [1,0,1]_{u} a^8 t^4
+ [1,0,1]_{\tilde{u}} a^8 t^4
\nn\\
&&
\hspace{0.2cm}
- [0,0,1]_{u} [1,0,0]_{\tilde{u}} a^{10} t^5
+ a^{16} t^8
)
\times
\PE\left[
[0,0,1]_{u} [1,0,0]_{\tilde{u}} a^2 t
\right]~,~
\nn\\
f_3
&=& 
\big(
1 
- [0,1,0]_{u} [0,1,0]_{\tilde{u}} a^4 t^2
+ ([1,0,0]_{u} [1,1,0]_{\tilde{u}}  + [0,1,1]_{u} [0,0,1]_{\tilde{u}} )a^6 t^3
\nn\\
&&
\hspace{0.2cm}
- ([0,1,2]_{u} + [1,0,1]_{u}[1,0,1]_{\tilde{u}}+ [2,1,0]_{\tilde{u}} )a^8 t^4
+ ([1,0,2]_{u} [1,0,0]_{\tilde{u}} 
\nn\\
&&
\hspace{0.2cm}
+ [0,0,1]_{u} [2,0,1]_{\tilde{u}} ) a^{10} t^5
+ ([2,0,0]_{u} [0,0,2]_{\tilde{u}}  - [0,0,2]_{u} [2,0,0]_{\tilde{u}}) a^{12} t^6
\nn\\
&&
\hspace{0.2cm}
- ([2,0,1]_{u}[0,0,1]_{\tilde{u}} + [1,0,0]_{u} [1,0,2]_{\tilde{u}}) a^{14} t^7
+ ([2,1,0]_{u} + [1,0,1]_{u} [1,0,1]_{\tilde{u}} 
\nn\\
&&
\hspace{0.2cm}
+ [0,1,2]_{\tilde{u}} ) a^{16} t^8
- ([1,1,0]_{u} [1,0,0]_{\tilde{u}} + [0,0,1]_{u} [0,1,1]_{\tilde{u}}) a^{18} t^9
\nn\\
&&
\hspace{0.2cm}
+ [0,1,0]_{u} [0,1,0]_{\tilde{u}} a^{20} t^{10}
- a^{24} t^{12}
\big)
\times
\PE\left[
[0,0,1]_{u} [1,0,0]_{\tilde{u}} a^2 t
\right]~.~
\eea

When one sums up the Hilbert series in \eref{es23a1b}, one obtains
\beal{es23a10}
&&
g(t,\tau,a,u,\tilde{u};\mathcal{M}_{U(3),4}) =
f_1 +
f_2 \left[
\frac{1}{1-\tau a^{-4} t}
+\frac{1}{1-\tau^{-1} a^{-4} t}
-2
\right]
\nn\\
&&
\hspace{6cm}
+ 
f_3
\frac{a^{-8} t^2}{(1- \tau a^{-4} t) (1- \tau^{-1} a^{-4} t)}
~.~
\eea
The Hilbert series has the following expansion up to order $t^4$,
\beal{es23a12b}
&&
g(t,\tau,a,u,\tilde{u};\mathcal{M}_{U(3),4}) =
1 
+ \big(
[0,0,1]_{u} [1,0,0]_{\tilde{u}} a^2
+ (\tau + \tau^{-1}) a^{-4}
\big) t
\nn\\
&&
\hspace{0.2cm}
+ \big(
([0,0,2]_{u} [2,0,0]_{\tilde{u}}) a^4
+ [0,0,1]_{u} [1,0,0]_{\tilde{u}} (\tau + \tau^{-1}) a^{-2}
+ (\tau^{2} + \tau^{-2} + 1) a^{-8}
\big) t^2
\nn\\
&&
\hspace{0.2cm}
+ \big(
(
[0,0,3]_{u} [3,0,0]_{\tilde{u}} 
+ [0,1,1]_{u} [1,1,0]_{\tilde{u}}
+ [1,0,0]_{u} [0,0,1]_{\tilde{u}}
) a^6
+ (
[0,0,2]_{u} [2,0,0]_{\tilde{u}}
\nn\\
&&
\hspace{0.7cm}
+ [0,1,0]_{u} [0,1,0]_{\tilde{u}}
) (\tau + \tau^{-1}) 
+[0,0,1]_{u} [1,0,0]_{\tilde{u}} (\tau^2 + \tau^{-2}) a^{-6}
\nn\\
&&
\hspace{0.7cm}
+ (\tau^3 + \tau + \tau^{-1} + \tau^{-3}) a^{-12}
\big) t^3
\nn\\
&&
\hspace{0.2cm}
+\big(
(
[0,0,4]_{u} [4,0,0]_{\tilde{u}} + [0,1,2]_{u} [2,1,0]_{\tilde{u}} + [0,2,0]_{u} [0,2,0]_{\tilde{u}} + [1,0,1]_{u} [1,0,1]_{\tilde{u}}
) a^8
\nn\\
&&
\hspace{0.2cm}
+ (
[0,0,3]_{u} [3,0,0]_{\tilde{u}} + [0,1,1]_{u} [1,1,0]_{\tilde{u}}
) (\tau + \tau^{-1}) a^2
+ (
[0,0,2]_{u} [2,0,0]_{\tilde{u}} 
\nn\\
&&
\hspace{0.7cm}
+ [0,1,0]_{u} [0,1,0]_{\tilde{u}}
) (\tau^2 + \tau^{-2}) a^{-4}
+ [0,0,1]_{u} [1,0,0]_{\tilde{u}} (\tau^3 + \tau^{-3}) a^{-10}
\nn\\
&&
\hspace{0.7cm}
+ (\tau^4 + \tau^2 + \tau^{-2} + \tau^{-4} + 1) a^{-16}
\big) t^4
+ \dots ~.~
\eea
The plethystic logarithm of the Hilbert series is
\beal{es23a12}
&&
\PL\left[
g(t,\tau,a,u,\tilde{u};\mathcal{M}_{U(3),4})
\right]
=
%16 p + 2 q
[0,0,1]_{u} [1,0,0]_{\tilde{u}} a^2 t
+ (\tau + \tau^{-1}) a^{-4} t
\nn\\
&&
\hspace{1.5cm}
%-p^4 - 32 p^3 q - 36 p^2 q^2
- a^8 t^4
- [1,0,0]_{u} [0,0,1]_{\tilde{u}} (\tau+\tau^{-1})  a^2 t^4
- [0,1,0]_{u} [0,1,0]_{\tilde{u}} a^{-4} t^4
+ \dots ~.~
\nn\\
\eea
The generators of the moduli space are identified from the plethystic logarithm as follows,
\beal{es23a13}
\ba{ccc}
\text{PL term} & \rightarrow & \text{generator}
\\
+ [0,0,1]_{u} [1,0,0]_{\tilde{u}} a^2 t  & \rightarrow & M_{i}^{j} = Q_{i}^{a} \tilde{Q}_{a}^{j}
\\
+ \tau a^{-4} t & \rightarrow & v_{+}
\\
+ \tau^{-1} a^{-4} t & \rightarrow & v_{-}
\ea
~~.~~
\eea
The generators form first order relations which are
\beal{es23a14}
\ba{ccc}
\text{PL term} & \rightarrow & \text{relation}
\\
- a^8 t^{4} & \rightarrow & \det{M} = 0
\\
- [1,0,0]_{u} [0,0,1]_{\tilde{u}} \tau  a^2 t^{4} & \rightarrow & 
v_{+} {R_{(3,4)}}_{i}^{j} = 0
\\
- [1,0,0]_{u} [0,0,1]_{\tilde{u}} \tau^{-1}  a^2 t^{4} & \rightarrow & 
v_{-} {R_{(3,4)}}_{i}^{j} = 0
\\
- [0,1,0]_{u} [0,1,0]_{\tilde{u}} a^{-4} t^{4} & \rightarrow & v_{+} v_{-} {R_{(2,4)}}_{i_1 i_2}^{j_1 j_2} = 0
\ea
~~,~~
\eea
where
\beal{es23a14b}
{R_{(3,4)}}_{i}^{j} &=& \frac{1}{6} \epsilon_{i k_1 k_2 k_3} \epsilon^{j m_1 m_2 m_3} M_{m_1}^{k_1} M_{m_2}^{k_2} M_{m_3}^{k_3} ~,~
\nn\\
{R_{(2,4)}}_{i_1 i_2}^{j_1 j_2} &=&
\frac{1}{2} \epsilon_{i_1 i_2 k_1 k_2} \epsilon^{j_1 j_2 m_1 m_2} M_{m_1}^{k_1} M_{m_2}^{k_2}
~.~
\eea

Given the generators and first order relations, the moduli space can be expressed as follows
\beal{es23a15}
\mathcal{M}_{U(3),4} = \mathbb{C}[M_{i}^{j},v_{\pm}] /
\langle
\det{M} = 0,
v_{\pm} {R_{(3,4)}}_{i}^{j} = 0 ,
v_{+} v_{-} {R_{(2,4)}}_{i_1 i_2}^{j_1 j_2} = 0
\rangle
~.~
\eea
Calling the space of $N\times N$ matrices with rank at most $k$ as $\mathcal{M}_{k,N}$,
the 4 components of the moduli space can be expressed as 
\beal{es23d1}
&
\mathcal{M}_{U(3),4}^{0} = 
\mathcal{M}_{3,4} ~,~
%\mathbb{C}[M_{i}^{j}] / \langle \det{M} = 0 \rangle ~,~
&
\nn\\
&
\mathcal{M}_{U(3),4}^{+} = 
\mathcal{M}_{2,4} \times \mathbb{C}[v_{+}]~,~
%\mathbb{C}[M_{i}^{j}] / \langle {R_{(3,4)}}^{i}_{j} = 0 \rangle  \times \mathbb{C}[v_{+}]~,~
\mathcal{M}_{U(3),4}^{-} = 
\mathcal{M}_{2,4} \times \mathbb{C}[v_{-}]~,~
%\mathbb{C}[M_{i}^{j}] / \langle {R_{(3,4)}}^{i}_{j} = 0 \rangle  \times \mathbb{C}[v_{-}]~,~
&
\nn\\
&
\mathcal{M}_{U(3),4}^{+-} =  
\mathcal{M}_{1,4} \times  \mathbb{C}[v_{+},v_{-}]~,~
%\mathbb{C}[M_{i}^{j}] / \langle {R_{(2,4)}}^{i}_{j} = 0 \rangle  \times  \mathbb{C}[v_{+},v_{-}]~,~
&
\eea
where $\mathcal{M}_{U(3),4}^{0}$ is the Higgs branch and $\mathcal{M}_{U(3),4}^{+}$, $\mathcal{M}_{U(3),4}^{-}$ and $\mathcal{M}_{U(3),4}^{+-}$ are mixed branches. There is no pure Coulomb branch for this theory. The corresponding Hilbert series are 
\beal{es23d2}
&&
g(t,\tau,a,u,\tilde{u}; \mathcal{M}_{U(3),4}^{0}) 
=
f_1
~,~
\nn\\
&&
g(t,\tau,a,u,\tilde{u}; \mathcal{M}_{U(3),4}^{+}) 
=
f_2 \times \frac{1}{1 - \tau a^{-4} t}
 ~,~
 \nn\\
 &&
g(t,\tau,a,u,\tilde{u}; \mathcal{M}_{U(3),4}^{-}) 
=
f_2 \times \frac{1}{1 - \tau^{-1} a^{-4} t}
 ~,~
 \nn\\
 &&
g(t,\tau,a,u,\tilde{u}; \mathcal{M}_{U(3),4}^{+-}) 
=
f_3 \times
\frac{1}{(1 - \tau a^{-4} t)(1 - \tau^{-1} a^{-4} t)}
 ~,~
\eea
where $f_1$, $f_2$ and $f_3$ are the dressing factors of the monopole operators in different GNO sublattices, as shown in \eref{es23a5b}.
The 4 components intersect in various subspaces which are
 \beal{es23d5}
& 
\mathcal{I}_{0} = \mathcal{M}_{1,4}
%\mathbb{C}[M_i^j] / \langle {R_{(2,4)}}_{i}^{j} = 0 \rangle
 ~,~
 \mathcal{I}_{M} = \mathcal{M}_{2,4}
 % \mathbb{C}[M_i^j] / \langle {R_{(3,4)}}_{i}^{j} = 0 \rangle 
 ~,~
 &
 \nn\\
 &
 \mathcal{I}_{+} = \mathcal{M}_{1,4} \times \mathbb{C}[v_{+}]
 %\mathbb{C}[M_i^j] / \langle {R_{(2,4)}}_{i}^{j} = 0 \rangle \times \mathbb{C}[v_{+}]
 ~,~
 \mathcal{I}_{-} = \mathcal{M}_{1,4} \times \mathbb{C}[v_{-}]
% \mathbb{C}[M_i^j] / \langle {R_{(2,4)}}_{i}^{j} = 0 \rangle \times \mathbb{C}[v_{-}]
 ~.~
 &
 \eea
The corresponding Hilbert series are
 \beal{es23d6}
 &
 g(t,\tau,a,u,\tilde{u};\mathcal{I}_{M} ) = f_2 ~,~
 g(t,\tau,a,u,\tilde{u}; \mathcal{I}_{0} ) = f_3 ~,~
&
\nn\\
&
g(t,\tau,a,u,\tilde{u}; \mathcal{I}_{+} ) =f_3 \times  \frac{1}{1 - \tau a^{-4} t} ~,~
g(t,\tau,a,u,\tilde{u}; \mathcal{I}_{-} ) = f_3 \times \frac{1}{1 - \tau^{-} a^{-4} t} ~.~
&
 \eea
 
The union of the 4 components is the moduli space.   
By removing contributions from the intersections, the Hilbert series of the full moduli space $\mathcal{M}_{U(3),4}$ can be expressed as
\beal{es23d10}
&&
g(t,\tau,a,u,\tilde{u}; \mathcal{M}_{U(3),4}) 
=
g(t,\tau,a,u,\tilde{u}; \mathcal{M}_{U(3),4}^{0}) + 
g(t,\tau,a,u,\tilde{u}; \mathcal{M}_{U(3),4}^{+}) 
\nn\\
&&
\hspace{2cm}
+
g(t,\tau,a,u,\tilde{u}; \mathcal{M}_{U(3),4}^{-}) +
g(t,\tau,a,u,\tilde{u}; \mathcal{M}_{U(3),4}^{+-})  
- 2 g(t,\tau,a,u,\tilde{u}; \mathcal{I}_{M} )
\nn\\
&&
\hspace{2cm}
- g(t,\tau,a,u,\tilde{u}; \mathcal{I}_{+} ) 
- g(t,\tau,a,u,\tilde{u}; \mathcal{I}_{-} )
+  g(t,\tau,a,u,\tilde{u}; \mathcal{I}_{0} )
~.~
\eea
This expression for the Hilbert series of the full moduli space is in agreement with the Hilbert series expression in \eref{es23a10}.
\\

%===============================%
\subsection{General Result of the Moduli Space \label{sgeneral}}

The Hilbert series $g(t,\tau,a,u,\tilde{u};\mathcal{M}_{U(N_c),N_f})$ which has been computed in the above sections all satisfy a general form. In order to present this general form, we make use of the \textit{highest weight generating function} for the characters of irreducible representations of the flavour symmetry $SU(N_f)_1 \times SU(N_f)_2$. The highest weight generating function for Hilbert series makes use of the map
\beal{es90c1}
\prod_{i=1}^{N_c -1} \mu_i^{n_i} \mapsto [n_1, \dots, n_{N_c - 1}]_{u}^{SU(N_f)_1} ~~,~~
\prod_{i=1}^{N_c -1} \nu_i^{n_i} \mapsto [n_1, \dots, n_{N_c - 1}]_{\tilde{u}}^{SU(N_f)_2} ~,~
\eea
where fugacities $\mu_i$ and $\nu_i$ count the highest weight of the irreducible representations of $SU(N_f)_1 \times SU(N_f)_2$.

Using the highest weight generating function of Hilbert series, one can for instance express concisely the dressing factor for monopole operators as follows,
\beal{es90c2}
\mathcal{F}_{N_c,N_f} = \PE\left[
\sum_{i=1}^{N_c} \mu_{N_f - i} \nu_{i} a^{2i} t^i
\right] ~.~
\eea
After the inclusion of the monopole operators the highest weight generating function is
\beal{es90c3}
&&
\mathcal{G}(t,\tau,a,u,\tilde{u}; \mathcal{M}_{U(N_c),N_f}) 
=
\mathcal{F}_{N_c,N_f} 
+ \mathcal{F}_{N_c-1, N_f} \left[
\frac{1}{1-\tau a^{-N_f} t} + 
\frac{1}{1-\tau^{-1} a^{-N_f} t} - 2
\right]
\nn\\
&&
\hspace{6cm}
+ \mathcal{F}_{N_c-2, N_f}
\frac{a^{-2N_f}t^2}{(1-\tau a^{-N_f} t) (1-\tau^{-1} a^{-N_f} t)}
~,~
\eea
where $t$ counts magnetic monopoles $v_{\pm}$ and mesonic operators $M_m=Q \tilde{Q}$ and corresponds to $U(1)$ symmetry which replaces $U(1)_R$. By identifying the exponents of fugacities $\mu_i$ and $\nu_i$ in the expansion of the highest weight generating function in \eref{es90c3}, one obtains the character expansion of the  Hilbert series.

The plethystic logarithm of the Hilbert series as a highest weight generating function is
\beal{es90a10}
%\PL\left[
%g(t,\tau,a,u,\tilde{u}; \mathcal{M}_{U(N_c),N_f}) 
%\right]
%&\mapsto&
&&
\mu_{N_f - 1} \nu_{1} 
a^2 t
+ 
(\tau + \tau^{-1}) a^{-N_f} t
\nn\\
&&
- \mu_{N_f - (N_c+1)} \nu_{N_c+1} a^{2(N_c+1)} t^{N_c + 1}
\nn\\
&&
- \mu_{N_f - N_c} \nu_{N_c} (\tau + \tau^{-1}) a^{2N_c - N_f} t^{N_c + 1}
\nn\\
&&
- \mu_{N_f - (N_c - 1)} \nu_{N_c - 1} a^{2 (N_c - 1) - 2 N_f} t^{N_c + 1}
+\dots ~.~
%\nn\\
\eea

The following product of mesonic operators is used in order to express relations amongst moduli space generators,
\beal{es90a11e1}
{R_{(N_c,N_f)}}_{i_1 \dots i_{N_f-N_c}}^{j_1 \dots j_{N_f-N_c}}
= 
\frac{1}{N_c !} 
\epsilon_{i_1 \dots i_{N_f - N_c} k_1 \dots k_{N_c}} 
\epsilon^{j_1 \dots j_{N_f - N_c} m_1 \dots m_{N_c}} 
M_{m_1}^{k_1} \dots 
M_{m_{N_c}}^{k_{N_c}} ~,~
\eea
where
\beal{es90a11e2}
R_{(N_c,N_c)} = \det (M) 
~.~
\eea
From the plethystic logarithm in \eref{es90a10}, the general form of the generators can be identified as
\beal{es90a15}
\ba{ccc}
\text{PL term} & \rightarrow & \text{generator}
\\
\mu_{N_f - 1} \nu_{1} a^2 t & \rightarrow & M_{i}^{j} = Q_{i}^{a} \tilde{Q}_{a}^{j}
\\
\tau^{\pm 1} a^{-N_f} t & \rightarrow & v_{\pm}
\ea
~~.~~
\eea
Furthermore, the general form of the first order relations formed amongst the generators are
\beal{es90a16}
\ba{ccc}
\text{PL term} & \rightarrow & \text{relation}
\\
- \mu_{N_f - (N_c+1)} \nu_{N_c+1} a^{2(N_c+1)} t^{N_c + 1}
 & \rightarrow &
 {R_{(N_c+1,N_f)}}_{i_1 \dots i_{N_f-N_c-1}}^{j_1 \dots j_{N_f-N_c-1}}
 %\text{minors}(N_c+1,M) 
 = 0
\\
- \mu_{N_f - N_c} \nu_{N_c} (\tau + \tau^{-1}) a^{2N_c - N_f} t^{N_c + 1}
& \rightarrow &
%v_\pm  \text{minors}(N_c,M) 
v_\pm {R_{(N_c,N_f)}}_{i_1 \dots i_{N_f-N_c}}^{j_1 \dots j_{N_f-N_c}} 
= 0
\\
- \mu_{N_f - (N_c - 1)} \nu_{N_c - 1} a^{2 (N_c - 1) - 2 N_f} t^{N_c + 1}
& \rightarrow &
%v_{+} v_{-} \text{minors}(N_c-1,M) 
v_{+} v_{-}  {R_{(N_c-1,N_f)}}_{i_1 \dots i_{N_f-N_c+1}}^{j_1 \dots j_{N_f-N_c+1}} 
= 0
\ea
~~.~~
\eea
It is important to note that the terms in the plethystic logarithm in \eref{es90a10} which correspond to the above relations do not appear in the Hilbert series expansion itself. This can be seen when one expands the dressing factor in \eref{es90c2} with the contributions from the monopole operators. One can show that the terms of the plethystic logarithm in \eref{es90a16} do not appear as operators in the Hilbert series expansion and that the relations in \eref{es90a16} are satisfied.

From the above analysis of the plethystic logarithm, the moduli space of the $U(N_c)$ theory with $N_f$ flavors can be expressed as the following algebraic variety,
\beal{es90a17}
\mathcal{M}_{U(N_c),N_f} =
\mathbb{C}[M_i^j,v_{\pm}] / \mathcal{I}~,~
\eea
where the quotienting ideal is
\beal{es90a20}
\mathcal{I} =
\langle
 {R_{(N_c+1,N_f)}} = 0 ~,~
v_{\pm}  {R_{(N_c,N_f)}} = 0 ~,~
v_{+}v_{-}  {R_{(N_c-1,N_f)}} = 0
\rangle
~.~
\eea 
Let us call $\mathcal{M}_{k,N}$ the space of all $N\times N$ matrices $M_i^j$ which at most have rank $k$. In terms of \eref{es90a11e1}, one can write $\mathcal{M}_{k,N}= \mathbb{C}[M_i^j]/ \langle R_{(k+1,N)} = 0 \rangle$. Then using $\mathcal{M}_{k,N}$, the 4 components of the moduli space can be expressed as
\beal{es90d1}
\mathcal{M}_{U(N_c),N_f}^{0} &=&
\mathcal{M}_{N_c,N_f} ~,~
%\mathbb{C}[M_{i}^{j}] / \langle {R_{(N_c+1,N_f)}}^{i}_{j} = 0 \rangle ~,~
\nn\\
\mathcal{M}_{U(N_c),N_f}^{+} &=& 
\mathcal{M}_{N_c-1,N_f} \times \mathbb{C}[v_{+}] ~,~
%\mathbb{C}[M_{i}^{j}] / \langle {R_{(N_c,N_f)}}^{i}_{j} = 0 \rangle  \times \mathbb{C}[v_{+}]~,~
\nn\\
\mathcal{M}_{U(N_c),N_f}^{-} &=& 
\mathcal{M}_{N_c-1,N_f} \times \mathbb{C}[v_{-}] ~,~
%\mathbb{C}[M_{i}^{j}] / \langle {R_{(N_c,N_f)}}^{i}_{j} = 0 \rangle  \times \mathbb{C}[v_{-}]~,~
\nn\\
\mathcal{M}_{U(N_c),N_f}^{+-} &=& 
\mathcal{M}_{N_c-2,N_f} \times \mathbb{C}[v_{+},v_{-}]~,~
%\mathbb{C}[M_{i}^{j}] / \langle {R_{(N_c-1,N_f)}}^{i}_{j} = 0 \rangle  \times \mathbb{C}[v_{+},v_{-}]~,~
\eea
where $\mathcal{M}_{U(N_c),N_f}^{0}$ is the Higgs branch, $\mathcal{M}_{U(N_c),N_f}^{+}$ and $\mathcal{M}_{U(N_c),N_f}^{-}$ are mixed branches, and $\mathcal{M}_{U(N_c),N_f}^{+-}$ is a Coulomb branch when $N_c=1,2$ and a mixed branch when $N_c>2$.\footnote{Note that component $\mathcal{M}_{U(N_c),N_f}^0$ is the dressing factor for components $\mathcal{M}_{U(N_c+1),N_f}^+$ and $\mathcal{M}_{U(N_c+1),N_f}^-$ and the dressing factor for component $\mathcal{M}_{U(N_c+2),N_f}^{+-}$.} The corresponding highest weight generating functions for the Hilbert series are
\beal{es90d2}
&&
\mathcal{G}(t,\tau,a,u,\tilde{u}; \mathcal{M}_{U(N_c),N_f}^{0}) 
=
\mathcal{F}_{N_c,N_f}
~,~
\nn\\
&&
\mathcal{G}(t,\tau,a,u,\tilde{u}; \mathcal{M}_{U(N_c),N_f}^{+}) 
=
\mathcal{F}_{N_c-1,N_f} \times \frac{1}{1 - \tau a^{-4} t}
 ~,~
 \nn\\
 &&
\mathcal{G}(t,\tau,a,u,\tilde{u}; \mathcal{M}_{U(N_c),N_f}^{-}) 
=
\mathcal{F}_{N_c-1,N_f} \times \frac{1}{1 - \tau^{-1} a^{-4} t}
 ~,~
 \nn\\
 &&
\mathcal{G}(t,\tau,a,u,\tilde{u}; \mathcal{M}_{U(N_c),N_f}^{+-}) 
=
\mathcal{F}_{N_c-2,N_f} \times
\frac{1}{(1 - \tau a^{-4} t)(1 - \tau^{-1} a^{-4} t)}
 ~,~
\eea
where $\mathcal{F}_{N_c,N_f}$, $\mathcal{F}_{N_c-1,N_f}$ and $\mathcal{F}_{N_c-2,N_f}$ are the dressing factors in \eref{es90c2} for the different GNO sublattices. The 4 components of the moduli space intersect in the following subspaces,
\beal{es90d3}
&
 \mathcal{I}_{M} = \mathcal{M}_{N_c-1,N_f}
 %\mathbb{C}[M_i^j] / \langle {R_{(N_c,N_f)}}_{i}^{j} = 0 \rangle 
 ~,~
   \mathcal{I}_{0} =  \mathcal{M}_{N_c-2,N_f}
  %\mathbb{C}[M_i^j] / \langle {R_{(N_c-1,N_f)}}_{i}^{j} = 0 \rangle 
 ~,~
 &\nn\\
 &
 \mathcal{I}_{+} = 
  \mathcal{M}_{N_c-2,N_f} \times \mathbb{C}[v_{+}]
 %\mathbb{C}[M_i^j] / \langle {R_{(N_c-1,N_f)}}_{i}^{j} = 0 \rangle  \times \mathbb{C}[v_{+}]
 ~,~
 \mathcal{I}_{-} = 
  \mathcal{M}_{N_c-2,N_f} \times \mathbb{C}[v_{-}]
 %\mathbb{C}[M_i^j] / \langle {R_{(N_c-1,N_f)}}_{i}^{j} = 0 \rangle  \times \mathbb{C}[v_{-}]
 ~.~
 &
 \eea
The corresponding highest weight generating functions for the Hilbert series are
 \beal{es90d4}
 &
\mathcal{G}(t,\tau,a,u,\tilde{u};\mathcal{I}_{M} ) = \mathcal{F}_{N_c-1,N_f} ~,~
\mathcal{G}(t,\tau,a,u,\tilde{u}; \mathcal{I}_{0} ) = \mathcal{F}_{N_c-2,N_f} ~,~
&
\nn\\
&
\mathcal{G}(t,\tau,a,u,\tilde{u}; \mathcal{I}_{+} ) =  \mathcal{F}_{N_c-2,N_f} \times \frac{1}{1 - \tau a^{-N_f} t} ~,~
\mathcal{G}(t,\tau,a,u,\tilde{u}; \mathcal{I}_{-} ) =  \mathcal{F}_{N_c-2,N_f} \times \frac{1}{1 - \tau^{-} a^{-N_f} t} ~.~
&
\nn\\
 \eea
Taking into account all the intersections, the highest weight generating function for the Hilbert series of the full moduli space $\mathcal{M}_{U(N_c),N_f}$ can be expressed as
\beal{es90d10}
&&
\mathcal{G}(t,\tau,a,u,\tilde{u}; \mathcal{M}_{U(N_c),N_f}) 
=
\mathcal{G}(t,\tau,a,u,\tilde{u}; \mathcal{M}_{U(N_c),N_f}^{0}) + 
\mathcal{G}(t,\tau,a,u,\tilde{u}; \mathcal{M}_{U(N_c),N_f}^{+}) 
\nn\\
&&
\hspace{2cm}
+
\mathcal{G}(t,\tau,a,u,\tilde{u}; \mathcal{M}_{U(N_c),N_f}^{-}) +
\mathcal{G}(t,\tau,a,u,\tilde{u}; \mathcal{M}_{U(N_c),N_f}^{+-})  
- 2 \mathcal{G}(t,\tau,a,u,\tilde{u}; \mathcal{I}_{M} )
\nn\\
&&
\hspace{2cm}
- \mathcal{G}(t,\tau,a,u,\tilde{u}; \mathcal{I}_{+} ) 
- \mathcal{G}(t,\tau,a,u,\tilde{u}; \mathcal{I}_{-} )
+  \mathcal{G}(t,\tau,a,u,\tilde{u}; \mathcal{I}_{0} )
~.~
\eea
This expression for the highest weight generating function for the Hilbert series of the full moduli space is in agreement with the Hilbert series expression in \eref{es90c3}.
\\

%==================================================%
%==================================================%
\section{The Superconformal Index and the Hilbert Series \label{scfi}}

In this section, we examine the relation between the superconformal index and the Hilbert series. The superconformal index by itself does not give information on the moduli space. Only by taking appropriate limits to a Hilbert series one can derive information about the structure of the moduli space. The following section proposes limits from the superconformal index which reproduce Hilbert series of certain subspaces of the moduli space of the 3d $\mathcal{N}=2$ theories. 
\\

%==================================================%
\subsection{The $\mathcal{N}=2$ Superconformal Index}

Firstly, let us recall the definition of the superconformal index for 3d $\mathcal N = 2$ theories. The bosonic subgroup of the 3d $\mathcal N = 2$ superconformal group is $SO(2,3) \times SO(2)$ whose three Cartan elements are denoted by $E,j$ and $R$. The superconformal index is defined by \cite{Bhattacharya:2008bja}
\begin{align}
I\left(x,\mathsf{u}_i\right) = \mathrm{Tr} (-1)^F \exp(-\beta'\{Q,S\}) x^{E+j} \left(\prod_i \mathsf u_i^{F_i}\right)
\end{align}
where $Q$ is a supercharge of quantum numbers $E = \frac{1}{2},j = -\frac{1}{2}$ and $R = 1$, and $S = Q^\dagger$. $x$ is the fugacity for $E+j$ and $\mathsf u_i$'s are additional fugacities for global symmetries of the theory. The trace is taken over the Hilbert space of the SCFT on $\mathbb R \times S^2$, or equivalently over the space of local gauge invariant operators on $\mathbb R^3$. As usual, only the BPS states, which saturate the inequality
\begin{align}
\{Q,S\} = E-R-j \geq 0,
\end{align}
contribute to the index. 

Using supersymmetric localization, the superconformal index can be exactly computed as follows, \cite{Kim:2009wb,Imamura:2011su}
\footnote
{
$(a;q)_n$ is the $q$-Pochhammer symbol, defined by
\begin{align}
(a;q)_n=\prod_{k=0}^{n-1} \left(1-a q^k\right).
\end{align}
}

\beal{eq:SCI}
&&
I(x,{}u,\tilde{{}u},{}a,\tau)
=
\nn\\
&&
\hspace{1cm}
\sum_{m\in\mathbb Z^{N_c}/S_{N_c}} \oint \left(\prod_{a = 1}^{N_c} \frac{dz_a}{2 \pi i z_a}\right) \frac{1}{|\mathcal W_m|} \tau^{\sum_a m_a} Z_\text{vector}(x,z,m) Z_\text{chiral}(x,{}u,\tilde{{}u},{}a,z,m)
~,~
\nn\\
\eea
where
\beal{IndexChiral}
Z_\text{vector}(x,z,m) &=&
\prod_{\substack{a,b = 1 \\ (a \neq b)}}^{N_c} x^{-|m_a-m_b|/2} \left(1-z_a z_b^{-1} x^{|m_a-m_b|}\right)
~,~
\nn\\
Z_\text{chiral}(x,{}u,\tilde{{}u},{}a,z,m)
&=&
\prod_{a = 1}^{N_c} x^{(1-r) N_f |m_a|} {}a^{-N_f |m_a|}
\nn\\
&&
\times \prod_{i = 1}^{N_f} \frac{\left(z_a^{-1} {}u_i^{-1} {}a^{-1} x^{|m_a|+2-r};x^2\right)_\infty \left(z_a \tilde{{}u}_i^{-1} {}a^{-1} x^{|m_a|+2-r};x^2\right)_\infty}{\left(z_a {}u_i {}a x^{|m_a|+r};x^2\right)_\infty \left(z_a^{-1} \tilde{{}u}_i {}a x^{|m_a|+r};x^2\right)_\infty}
~.~
\nn\\
\eea
Above, $|\mathcal W_m|$ is the Weyl group order of the residual gauge group left unbroken by flux $m$. We have taken into account the gauge group $U(N_c)$ and the matter content: the $N_f$ pairs of fundamental and anti-fundamental chiral multiplets. ${}u = ({}u_1,\ldots,{}u_{N_f}),\tilde{{}u} = (\tilde{{}u}_1,\ldots,\tilde{{}u}_{N_f}),{}a$ and $\tau$ are the fugacities for the global symmetry $SU(N_f)_1 \times SU(N_f)_2 \times U(1)_A \times U(1)_T$ respectively. Note that $\prod_{i = 1}^{N_f} {}u_i = \prod_{i = 1}^{N_f} \tilde{{}u}_i = 1$.
\\

%---------------------------------%
\subsection{Limits of the $\mathcal{N}=4$ Superconformal Index}

Let us review the proposal \cite{Razamat:2014pta} for the relation between the superconformal index and the Hilbert series of $\mathcal N = 4$ theories \cite{Benvenuti:2010pq,Cremonesi:2013lqa}.  Let us denote by $j_H$ and $j_V$ the spins of the two $SU(2)$ in the $SO(4)_R = SU(2)_H \times SU(2)_V$ $R$-symmetry.  $x$ is the $E+j$ fugacity and $x'$ is the $j_H-j_V$ fugacity. The superconformal index for an $\mathcal N = 4$ theory is 
\begin{align}\label{esindex2}
\begin{aligned}
I(x,x') &= \mathrm {Tr}' (-1)^F x^{E+j} x'^{j_H-j_V} \\
&= \mathrm {Tr}' (-1)^F t_H^{E-j_V} t_C^{E-j_H}
\end{aligned}
\end{align}
where we ignore other global symmetry fugacities and $t_H = x x'$, $t_C= x x'^{-1}$. The primed trace $\mathrm{Tr}'$ denotes that the trace is taken over the BPS states. The BPS condition $E = j_H+j_V+j$ \cite{2010NuPhB.827..183I} is used for the second equality. Under $\mathcal{N}=2$ twisting some of the fermions in the $\mathcal{N}=4$ vector multiplet get the same quantum numbers as the F-terms and play the same role for the index as the F-terms for the Hilbert series. It is important to note that the index is unreliable when there are accidental IR corrections to the R-symmetry.

The proposed limits for getting the Hilbert series of the Higgs branch and the Coulomb branch from the superconformal index are \footnote{A crucial comment here is that the Higgs branch limit gives the Hilbert series only when a complete Higgsing of the gauge group occurs along the Higgs branch.}
\begin{align} \label{eq:N=4 limits}
\begin{array}{lll}
\text{Higgs branch:} &\mathrm{HS}_H(t_H) = \lim_{t_C \rightarrow 0} I(t_H,t_C)~,~ \\
\text{Coulomb branch:} &\mathrm{HS}_C(t_C) = \lim_{t_H \rightarrow 0} I(t_H,t_C)~,~
\end{array}
\end{align}
where $\mathrm{HS}_H$ and $\mathrm{HS}_C$ are respectively the Hilbert series of the Higgs and Coulomb branches.

Note that the BPS condition $E = j_H+j_V+j$ implies inequalities $E \geq j_H$ and $E \geq j_V$. Using \eref{esindex2} the first limit in \eref{eq:N=4 limits} restricts to BPS states with $E = j_{H}$ implying $j_{V} = j = 0$. Similar arguments apply for the second limit in \eref{eq:N=4 limits}. Therefore, the index in each limit captures the $SU(2)_{V/H}$ singlet scalar BPS states, which corresponds to the Hilbert series of the Higgs/Coulomb branch of the $\mathcal N = 4$ theory, respectively.
\\

%==================================================%
\subsection{Generalized Limits for the $\mathcal{N}=2$ Superconformal Index}

$\mathcal N = 2$ theories do not in general have distinct Higgs and Coulomb branches. Furthermore, there is only one $U(1)_R$ symmetry in the superconformal algebra for $\mathcal N=2$ theories. Nevertheless, one may try to generalize the limits in \eqref{eq:N=4 limits} for $\mathcal N = 2$ theories. The $\mathcal{N}=2$ $U(1)_R$ charge plays the role of $j_H + j_V$ in $\mathcal{N}=4$. In addition, one can choose one of the $\mathcal{N}=2$ global $U(1)$ symmetries and choose its charge to play the role of $j_H-j_V$ in $\mathcal{N}=4$. With these choices, it turns out that the resulting generalized limits of the $\mathcal{N}=2$ superconformal index give rise to Hilbert series of certain subspaces of the moduli space for the $\mathcal{N}=2$ theory. In addition, such generalized limits of the $\mathcal{N}=2$ superconformal index are not unique because the $\mathcal N = 2$ theories we are considering have several $U(1)$ global symmetries.

We will examine 4 limits of the superconformal index of the $\mathcal N = 2$ $U(N_c)$ theory with $N_f$ flavors. The BPS condition and certain constraints on the global $U(1)$ symmetry charges, which derive from the requirement that the limit is well-defined and non-divergent, can be used to show that there are just 4 relevant limits to consider. This is further elaborated in the following section. Here it is noted that each of the 4 limits corresponds to a Hilbert series of a certain subspace of the moduli space. 3 of them can be expressed in terms of the 4 main components of the moduli space which are discussed in section \sref{sgeneral}. These 3 subspaces are as follows:
\begin{itemize}
\item $\mathcal{M}_{U(N_c),N_f}/\left<v_\pm = 0\right> = \mathcal{M}_{U(N_c),N_f}^{0}$

\item $\mathcal{M}_{U(N_c),N_f}/\left<v_- = 0\right> = \mathcal{M}_{U(N_c),N_f}^{0} \cup \mathcal{M}_{U(N_c),N_f}^{+}$

\item $\mathcal{M}_{U(N_c),N_f}/\left<v_+ = 0\right> = \mathcal{M}_{U(N_c),N_f}^{0} \cup \mathcal{M}_{U(N_c),N_f}^{-}$
\end{itemize}
The 4th limit gives the Hilbert series of a subspace of the moduli space that cannot be directly expressed in terms of the 4 main components. It is a subspace of component $\mathcal{M}_{U(N_c),N_f}^{+-}$ as follows:
\begin{itemize}
\item $\mathcal{M}_{U(N_c),N_f}/\left<M_i^j = 0\right>  \subset \mathcal{M}_{U(N_c),N_f}^{+-}$
\end{itemize}
 By considering all 4 limits, we are going to see that taking a limit of the superconformal index cannot reproduce the Hilbert series of the whole component $\mathcal{M}_{U(N_c),N_f}^{+-}$, and thus that of the complete moduli space. The subsequent sections explain how we obtain the Hilbert series of each subspace from the superconformal index.
\\

\comment{ The superconformal index captures both bosonic contributions and fermionic contributions. Since the fermionic contributions are accompanied by the factor $(-1)$, there are cancelations between the bosonic and the fermionic contributions. They might reflect the transition along the RG-flow from a BPS boson-fermion pair with appropriate quantum numbers to a non-BPS state, but they also might be accidental; i.e., both the boson and the fermion remain BPS but their contributions just cancel out each other.
%\footnote{Bashkirov} 
In principle, because of such accidental cancelations, the superconformal index is not a convenient observable to deal with if we are only interested in the chiral ring or the moduli space of the theory as we do here.

In general, we believe that such a transition from a BPS pair to a non-BPS state does occur whenever it can. Thus, one might expect that a cancelation in the index indicates the transition such that a nontrivial chiral ring element should leave its contribution in the index as a non-vanishing term. However, that is true only when the global symmetry of the theory is Abelian. If the global symmetry is non-Abelian, the index is written in terms of various representation characters of the global symmetry. In that case, the transition would occur only when the whole character is canceled out. A partial cancelation between different representation characters doesn't correspond to the transition. Therefore, one cannot avoid the accidental cancelations in the superconformal index.
}

Recall why the limits in \eqref{eq:N=4 limits} capture scalar BPS states: if energy $E$ of a BPS state is equal to the $R$-charge $j_{H/V}$, the state is scalar BPS due to the $\mathcal N = 4$ BPS condition $E \geq j_H+j_V+j$ \cite{2010NuPhB.827..183I}. The idea for $\mathcal N = 2$ theories is the same. We try to identify a state whose energy is equal to the $U(1)_R$ charge $R$. Such a state then should be scalar BPS because of the $\mathcal N=2$ BPS condition $E \geq R+j$. We cannot trace every scalar BPS state by taking a limit of the superconformal index because there are accidental cancelations between the bosonic and the fermionic contributions to the index. This section explains which remaining states can be traced by taking an appropriate limit of the $\mathcal{N}=2$ superconformal index.

For every factor $U(1)_k$ in the global symmetry of the theory, one can introduce a corresponding fugacity $\mathsf u_k$.
In order to have a well-defined non-divergent limit of the superconformal index, we propose the condition that for a $U(1)_k$ factor in the global symmetry, the ratio of the $U(1)_k$ charge $F_k$ to the $U(1)_R$ charge $R$ satisfies the following bound
\begin{align} \label{eq:F/R}
\frac{F_k}{R} \leq 1 ~.~
\end{align}
We have assumed for simplicity that $F_k$ is normalized such that the right hand side is 1. The role of the above condition is going to become clearer when one revisits the general form of the $\mathcal{N}=2$ index
\begin{align*}
I\left(x,\mathsf u_i\right) = \mathrm{Tr}' (-1)^F x^{E+j} \left(\prod_i \mathsf u_i^{F_i}\right) ~,~
\end{align*}
where we can make shifts of the $E+j$ fugacity and the $U(1)_k$ fugacity, $x \rightarrow x y,\mathsf u_k \rightarrow \mathsf u_k y^{-1}$, such that in the limit $y \rightarrow 0$ one has
\begin{align}
\lim_{y \rightarrow 0} I(x y,\mathsf u_{i (\neq k)},\mathsf u_k y^{-1}) = \lim_{y \rightarrow 0} \mathrm{Tr}' (-1)^F x^{2 E-R} \left(\prod_i \mathsf u_i^{F_i}\right) y^{2 (E-R)+R-F_k}.
\end{align}
Again the primed trace $\mathrm{Tr}'$ denotes that the trace is taken over the BPS states. Given the BPS condition $E \geq R+j$ and the condition $R \geq F_k$ from \eref{eq:F/R}, the power of $y$ for each term is non-negative. Therefore, the limit $y \rightarrow 0$ only leaves terms which are independent of $y$. The remaining terms correspond to the contributions of BPS states satisfying $E = R = F_k$ and $j = 0$. This is exactly the Hilbert series counting scalar BPS states of the theory,
\begin{align} \label{eq:generalized limit}
g(x,\mathsf u_i) = \lim_{y \rightarrow 0} I(x y,\mathsf u_{i (\neq k)},\mathsf u_k y^{-1}) ~,~
\end{align}
where $\mathsf u_i$'s are the global symmetry fugacities and $x$ is the energy fugacity.

The choice of global $U(1)$ symmetries, the constraints set by the $\mathcal{N}=2$ BPS condition, and the requirement for having a well-defined non-divergent limit of the superconformal index all lead to precisely 4 limits of the $\mathcal{N}=2$ superconformal index for the theories we are considering. In the following sections, these limits are presented and the resulting Hilbert series are identified with subspaces of the moduli space of the $\mathcal{N}=2$ theory.
\\

\comment{
Note that the $\mathcal N = 4$ results \eqref{eq:N=4 limits} are nothing but the special cases of taking $\mathsf u_k = x'$ or $\tilde {\mathsf u}_k = x'^{-1}$. For each $\mathcal{N}=4$ limit in, the condition \eqref{eq:F/R} corresponds to the condition $j_V \geq 0$ or $j_H \geq 0$, which is obviously satisfied. Here, $F_k = j_H-j_V$ or $F_k = j_V-j_H$ for each case, and $R = j_H+j_V$. As a result, the first limit in \eqref{eq:N=4 limits} gives rise to the Higgs branch Hilbert series generated by the mesonic generators, which satisfy $R-F_k = 2 R_V = 0$, whereas the second limit gives rise to the Coulomb branch Hilbert series generated by the monopole generators, which satisfy $R-F_k = 2 R_H = 0$. On the other hand, for general $\mathcal N = 2$ theories, the condition \eqref{eq:F/R} is not easy to check because the global charge doesn't have a fundamental restriction. Nevertheless, if we assume that a BPS boson-fermion pair becomes a non-BPS state along the RG-flow whenever it can and that no accidental IR symmetry emerges, it is enough to check the global charges of the UV BPS states. Thus, with those assumptions, we can show that some $U(1)$ global symmetries satisfy the condition \eqref{eq:F/R} in our case such that we have nontrivial limits of the superconformal index reproducing the Hilbert series of certain subspaces of the moduli space.
\\
}

%==================================================%
\subsubsection{$\mathcal{M}_{U(N_c),N_f}/\left<v_\pm = 0\right>$ and $\mathcal{M}_{U(N_c),N_f}/\left<M_i^j = 0\right>$} \label{sec:boundary I/III}

We are considering $\mathcal N = 2$ $U(N_c)$ theories with $N_f$ pairs of fundamental and anti-fundamental chiral multiplets which have a global symmetry of $SU(N_f)_1 \times SU(N_f)_2 \times U(1)_A \times U(1)_T$. Let us consider here the $U(1)_A$ axial symmetry. Given the charge assignments summarized in \tref{tfields}, one can identify bounds for the ratio of the $U(1)_A$ charge $A$ to the $U(1)_R$ charge $R$ for a BPS state as follows:
\begin{align} \label{eq:A/R}
-\frac{N_f}{(1-r) N_f-N_c+1} \leq \frac{A}{R} \leq \frac{1}{r}
\end{align}
where $r$ is the $U(1)_R$ charge of the fundamental and anti-fundamental chiral multiplets $Q$ and $\tilde Q$. $r$ is such that $\Delta_M = 2 r$ and $\Delta_V = (1-r) N_f-N_c+1$ for mesonic and monopole operators respectively are larger than or equal to 1/2 due to unitarity. We can take two differently normalized versions of $U(1)_A$ such that each inequality in \eqref{eq:A/R} takes the form of \eqref{eq:F/R}. Then, as we have proposed, the Hilbert series of a subspace of the moduli space generated by generators saturating each inequality can be obtained from the superconformal index. It turns out that the right inequality is saturated for the mesonic operators $M_i^j$, which have the $U(1)_A$ charge 2 and the $U(1)_R$ charge $2 r$, whereas the left inequality is saturated for the monopole operators $v_\pm$, which have the $U(1)_A$ charge $-N_f$ and the $U(1)_R$ charge $(1-r) N_f-N_c+1$. Therefore, we propose two limits of the superconformal index which give rise to the Hilbert series of 
two subspaces of the moduli space $\mathcal M_{U(N_c),N_f}/\left<v_\pm=0\right>$ and $\mathcal M_{U(N_c),N_f}/\left<M_i^j=0\right>$. $\mathcal M_{U(N_c),N_f}/\left<v_\pm=0\right>$ is the same as component $\mathcal{M}_{U(N_c),N_f}^{0}$ of the moduli space as discussed in section \ref{sgeneral} while $\mathcal M_{U(N_c),N_f}/\left<M_i^j=0\right>$ is only a subspace of component $\mathcal{M}_{U(N_c),N_f}^{+-}$:
\begin{gather}
\begin{gathered}
\mathcal M_{U(N_c),N_f}/\left<v_\pm = 0\right> = \mathcal M_{U(N_c),N_f}^{0}, \\
\mathcal M_{U(N_c),N_f}/\left<M_i^j = 0\right> = \mathcal M_{U(N_c),N_f}^{+-}/\left<M_i^j = 0\right> \subset \mathcal M_{U(N_c),N_f}^{+-}.
\end{gathered}
\end{gather}
Their Hilbert series are given by
\begin{align}
g(x,\tau,{}a,{}u,\tilde{{}u},;\mathcal M_{U(N_c),N_f}/\left<v_\pm=0\right>) &= \lim_{y \rightarrow 0} I(x y,{}u,\tilde{{}u},{}a y^{-r},\tau), \label{eq:N=2 limit1}\\
g(x,\tau,{}a,{}u,\tilde{{}u};\mathcal M_{U(N_c),N_f}/\left<M_i^j=0\right>) &= \lim_{y \rightarrow 0} I(x y,{}u,\tilde{{}u},{}a y^{(1-r)-(N_c-1)/N_f},\tau)~.~ 
\label{eq:N=2 limit2}
\end{align}
Again $x$ is the energy fugacity of the Hilbert series. ${}u,\tilde{{}u},{}a$ and $\tau$ are identified as the fugacities for $SU(N_f)_1 \times SU(N_f)_2 \times U(1)_A \times U(1)_T$ respectively.
\\

\paragraph{Computation.} Using the limits, we claim that one can obtain the explicit formulae for the Hilbert series of the two subspaces $\mathcal M_{U(N_c),N_f}/\left<v_\pm=0\right>$ and $\mathcal M_{U(N_c),N_f}/\left<M_i^j=0\right>$ from the superconformal index. Firstly, the Hilbert series of $\mathcal M_{U(N_c),N_f}/\left<v_\pm=0\right>$ is given by the limit \eqref{eq:N=2 limit1}. Since $\mathcal M_{U(N_c),N_f}/\left<v_\pm=0\right>$ is the same as component $\mathcal{M}_{U(N_c),N_f}^0$ of the moduli space,
\begin{align}
g(\mathcal M_{U(N_c),N_f}/\left<v_\pm=0\right>)=g(\mathcal{M}^{0}_{U(N_c),N_f}).
\end{align}
In this limit the monomial factor $x^{(1-r) N_f \sum_a |m_a|-\sum_{a<b} |m_a-m_b|} {}a^{-N_f \sum_a |m_a|}$ of the integrand in \eqref{eq:SCI} vanishes unless $m = \vec 0$. This is because the power of $x$, which is equal to $\Delta(m) = (1-r) N_f \sum_a |m_a|-\sum_{a < b} |m_a-m_b|$, should be positive for nonzero $m$. Therefore, only the $m = \vec 0$ contribution remains such that
\beal{eq:component1}
&&
g(x,a,{}u,\tilde{{}u};\mathcal{M}^{0}_{U(N_c),N_f}) =
 \lim_{y \rightarrow 0} I(x y,{}u,\tilde{{}u},{}a y^{-r},\tau) \nonumber
 \nn\\
&&
\hspace{1.5cm}
= \oint \ud\mu_{U(N_c)} \prod_{a = 1}^{N_c} \prod_{i = 1}^{N_f} \frac{1}{\left(1-z_a {}u_i a x^r\right) \left(1-z_a^{-1} \tilde{{}u}_i a x^r\right)} \label{eq:comp1formula}
~,~
\eea
where $d\mu_{U(N_c)}$ is the Haar measure for $U(N_c)$. The formula in \eqref{eq:comp1formula}, which is obtained from the index formula \eqref{eq:SCI}, is equivalent to the classical contribution of the mesonic operators in \eqref{es0a26} if we substitute $x = \textbf t$.

Next the limit \eqref{eq:N=2 limit2} gives the Hilbert series of $\mathcal M_{U(N_c),N_f}/\left<M_i^j=0\right>$. We consider the $U(1)$ case first and then consider general $U(N_c)$ cases with $N_c \geq 2$. For a $U(1)$ theory the vector multiplet does not contribute to the index. Only the contribution of chiral multiplets is nontrivial, which becomes the monomial factor
%%%%%%%
\begin{align}
x^{(1-r)N_f|m|} {}a^{-N_f |m|}
\end{align}
%%%%%%%
under the limit. For the $U(1)$ theory, $\mathcal M_{U(1),N_f}/\left<M_i^j=0\right>$ is nothing but component $\mathcal{M}_{U(1),N_f}^{+-}$ of the moduli space. Therefore,
\begin{align}
g(\mathcal M_{U(1),N_f}/\left<M_i^j=0\right>) = g(\mathcal{M}^{+-}_{U(1),N_f})
\end{align}
and
\beal{eq:CoulombU(1)}
g(x,\tau,a;\mathcal{M}^{+-}_{U(1),N_f})
&=& \lim_{y \rightarrow 0} I(x y,{}u,\tilde{{}u},{}a y^{1-r},\tau)
\nn\\
&=& \sum_{m =-\infty}^\infty \tau^{m} a^{-N_f |m|} x^{(1-r) N_f |m|} \nn\\
&=& \frac{1-a^{-2 N_f} x^{2 (1-r) N_f}}{(1-\tau a^{-N_f} x^{(1-r) N_f}) \left(1-\tau^{-1} a^{-N_f} x^{(1-r) N_f}\right)}
~.~
\eea
For a $U(1)$ theory, the nontrivial components of the moduli space are only component $\mathcal{M}_{U(1),N_f}^{0}$ and $\mathcal{M}_{U(1),N_f}^{+-}$ because component $\mathcal{M}_{U(1),N_f}^{+}$ and $\mathcal{M}_{U(1),N_f}^{-}$ are included in $\mathcal{M}_{U(1),N_f}^{+-}$. The Hilbert series of component $\mathcal{M}_{U(1),N_f}^{0}$ is given by \eqref{eq:component1} and the Hilbert series of component $\mathcal{M}_{U(1),N_f}^{+-}$ is given by \eqref{eq:CoulombU(1)}. Taking into account the fact that their intersection is only the origin, for this special case of the $U(1)$ theory, the complete Hilbert series can be written as
\beal{exx100}
&&
g(x,\tau,a,{}u,\tilde{{}u},\tau;\mathcal{M}_{U(1),N_f})
= g(x,a,{}u,\tilde{{}u};\mathcal{M}^{0}_{U(1),N_f})+g(x,\tau,a;\mathcal{M}^{+-}_{U(1),N_f})
\nn\\
&&
\hspace{6cm}
-g(\mathcal{M}^{0}_{U(1),N_f} \cap \mathcal{M}^{+-}_{U(1),N_f})
~,~
\nn\\
\eea
where we use
\begin{align}
g(\mathcal{M}^{0}_{U(1),N_f} \cap \mathcal{M}^{+-}_{U(1),N_f}) = 1
~.~
\end{align}
If we substitute $x = \textbf t$ into \eqref{exx100}, we recover the result in section \ref{s2}.

Now let us consider a $U(N_c)$ theory with $N_c \geq 2$. In this case, the superconformal index in the limit \eqref{eq:N=2 limit2} is given by
\begin{align}
\begin{aligned} \label{eq:component3}
g(x,\tau,a;\mathcal M_{U(N_c \geq 2),N_f}/\left<M_i^j=0\right>)
&= \lim_{y \rightarrow 0} I(x y,{}u,\tilde{{}u},{}a y^{(1-r)-(N_c-1)/N_f},\tau) \\
&= \sum_{m_1 = 0}^\infty \sum_{m_{2} = -\infty }^{0} \tau^{m_1+m_{2}} a^{-N_f (m_1-m_{2})} x^{(1-r) N_f (m_1-m_{2})} \\
&= \frac{1}{(1-\tau a^{-N_f} x^{(1-r) N_f-N_c+1}) \left(1-\tau^{-1} a^{-N_f} x^{(1-r) N_f-N_c+1}\right)}.
\end{aligned}
\end{align}
The above Hilbert series shows that the chiral ring is freely generated by two monopole operators $v_\pm$. Note that especially for $N_c = 2$, $\mathcal M_{U(2),N_f}/\left<M_i^j = 0\right>$ is again component $\mathcal{M}_{U(2),N_f}^{+-}$. Therefore,
\begin{align}
g(\mathcal M_{U(2),N_f}/\left<M_i^j=0\right>) = g(\mathcal{M}^{+-}_{U(2),N_f})
~.~
\end{align}
\\

%==================================================%
\subsubsection{$\mathcal{M}_{U(N_c),N_f}/\left<v_- = 0\right>$ and $\mathcal{M}_{U(N_c),N_f}/\left<v_+ = 0\right>$}

Let us consider in this section the topological symmetry $U(1)_T$. Given that only monopole operators are charged under $U(1)_T$, we do not directly use the $U(1)_T$ symmetry for formulating the $\mathcal{N}=2$ limit but use mixed symmetries $U(1)_+$ and $U(1)_-$ instead whose conserved currents are defined by
\begin{align}
J_\pm = r J_A\pm(N_f-N_c+1) J_T
\end{align}
where $J_A$ and $J_T$ are the conserved currents of $U(1)_A$ and $U(1)_T$. Following \tref{tfields}, one can show that the ratios of $U(1)_\pm$ to the $R$-charge are bounded from above as follows,
\beal{esxx100}
\frac{F_+}{R} &\leq& 1, \label{eq:F+/R} 
~,~
\\
\frac{F_-}{R} &\leq& 1 \label{eq:F-/R}
~,~
\eea
where $F_{\pm}$ are charges under $U(1)_{\pm}$.

\begin{table}
\centering
\begin{tabular}{c|cc}
\; & $F_{+}$ & $F_{-}$
\\
\hline
$v_+$ & $(1-r) N_f-N_c+1$ & $-(1+r) N_f+N_c-1$ 
\\
$v_-$ &  $-(1+r) N_f+N_c-1$ & $(1-r) N_f-N_c+1$
\end{tabular}
\caption{Saturated $U(1)_+$ and $U(1)_-$ charges for the monopole operators $v_\pm$. \label{tchargesF}}
\end{table}

Recall that only the monopole operators $v_\pm$ are charged under $U(1)_T$ with the charges $T = \pm1$. Thus, the mesonic operators $M_i^j$ just have the $U(1)_\pm$ charges $F_\pm = r A = 2 r$ and saturate both inequalities \eqref{eq:F+/R} and \eqref{eq:F-/R}. On the other hand, the two monopole operators have different $U(1)_\pm$ charges and are summarized in \tref{tchargesF}. As a result, $v_+$ saturates the bound \eqref{eq:F+/R} while $v_-$ saturates the bound \eqref{eq:F-/R}. Furthermore, the inequality \eqref{eq:F+/R} is saturated at a subspace of the moduli space $\mathcal {M}_{U(N_c),N_f}/\left<v_- = 0\right>$ while the inequality \eqref{eq:F-/R} is saturated at a subspace of the moduli space $\mathcal{M}_{U(N_c),N_f}/\left<v_+ = 0\right>$. Each subspace can be expressed in terms of the main components of the moduli space
\begin{align}
\begin{aligned}
\mathcal{M}_{U(N_c),N_f}/\left<v_- = 0\right> &= \mathcal{M}_{U(N_c),N_f}^{0} \cup \mathcal{M}^{+}_{U(N_c),N_f}, \\
\mathcal{M}_{U(N_c),N_f}/\left<v_+ = 0\right> &= \mathcal{M}_{U(N_c),N_f}^{0} \cup \mathcal{M}_{U(N_c),N_f}^{-}.
\end{aligned}
\end{align}
\\

\paragraph{Computation.} In order to obtain the Hilbert series of $\mathcal M/\left<v_- = 0\right>$, we propose the following limit of the superconformal index,
\begin{align}\label{elimitx1}
\begin{aligned} 
g(x,u_+;\mathcal{M}_{U(N_c),N_f}/\left<v_- = 0\right>) &= \lim_{y \rightarrow 0} I(x y,u_+ y^{-1}) \\
&=\lim_{y\rightarrow 0} \mathrm{Tr}' (-1)^F (x y)^{E+j} (u_+ y^{-1})^{r A+(N_f-N_c+1) T} \\
&=\lim_{y\rightarrow 0} \mathrm{Tr}' (-1)^F x^{2 E-R} u_+^{r A+(N_f-N_c+1) T} y^{2 E-R-r A-(N_f-N_c+1) T}
~,~
\end{aligned}
\end{align}
where $u_+$ is the fugacity of $U(1)_+$ and the other global symmetry fugacities are omitted. This tells us that only the contributions satisfying $E = R = r A+(N_f-N_c+1) T$ and $j = 0$ remain under the limit. One can check that the shift $u_+ \rightarrow u_+ y^{-1}$ here is equivalent to the shifts of the $U(1)_A$ and $U(1)_T$ fugacities ${}a \rightarrow {}a y^{-r}$ and $\tau \rightarrow \tau y^{-(N_f-N_c+1)}$ respectively. This is because $u_+$ and $u_-$ are written in terms of ${}a,\tau$ as $u_\pm = {}a^\frac{1}{2 r} \tau^{\pm\frac{1}{2 (N_f-N_c+1)}}$. Therefore, \eref{elimitx1} takes the form
\begin{align}
\begin{aligned} \label{eq:component+}
&g(x,\tau,a,{}u,\tilde{{}u},\tau;\mathcal{M}_{U(N_c),N_f}/\left<v_- = 0\right>) \\
&= \lim_{y \rightarrow 0} I(x y,{}u,\tilde{{}u},{}a y^{-r},\tau y^{-(N_f-N_c+1)}) \\
&= \oint \ud\mu_{U(N_c)} \prod_{a = 1}^{N_c} \prod_{i = 1}^{N_f} \frac{1}{\left(1-z_a {}u_i a x^r\right) \left(1- z_a^{-1} \tilde{{}u}_i a x^r\right)} \\
&\qquad +\sum_{m_1 = 1}^\infty \tau^{m_1} a^{-N_f m_1} x^{(1-r) N_f m_1} \oint \ud\mu_{U(N_c-1)} \prod_{a = 1}^{N_c-1} \prod_{i = 1}^{N_f} \frac{1}{\left(1-z_a {}u_i a x^r\right) \left(1-z_a^{-1} \tilde{{}u}_i a x^r\right)} \\
&= g(x,a,{}u,\tilde{{}u};\mathcal{M}^{0}_{U(N_c),N_f})+\frac{\tau a^{-N_f} x^{(1-r) N_f-N_c+1}}{1-\tau a^{-N_f} x^{(1-r) N_f-N_c+1}} \times g(x,a,{}u,\tilde{{}u};\mathcal{M}^{0}_{U(N_c-1),N_f}).
\end{aligned}
\end{align}
This is the same as the Hilbert series of the union of components $\mathcal{M}_{U(N_c),N_f}^{0}$ and $\mathcal{M}_{U(N_c),N_f}^{+}$,
\begin{align}
g(\mathcal M_{U(N_c),N_f}/\left<v_- = 0\right>) = g(\mathcal{M}^{0}_{U(N_c),N_f})+g(\mathcal{M}^{+}_{U(N_c),N_f})-g(\mathcal{M}^{0}_{U(N_c),N_f}\cap \mathcal{M}^{+}_{U(N_c),N_f})
~,~
\end{align}
where 
\beal{esx2000}
\mathcal{M}^{0}_{U(N_c),N_f}\cap \mathcal{M}^{+}_{U(N_c),N_f}
=
\mathcal{M}^{0}_{U(N_c-1),N_f} ~.~
\eea

In the same way, the Hilbert series of $\mathcal{M}_{U(N_c),N_f}/\left<v_+ = 0\right>$ is obtained from the superconformal index as follows:
\begin{align} \label{eq:component-}
\begin{aligned}
&g(x,\tau,a,{}u,\tilde{{}u},\tau;\mathcal{M}_{U(N_c),N_f}/\left<v_+ = 0\right>) \\
&= \lim_{y \rightarrow 0} I(x y,{}u,\tilde{{}u},{}a y^{-r},\tau y^{N_f-N_c+1}) \\
&= \oint \ud\mu_{U(N_c)} \prod_{a = 1}^{N_c} \prod_{i = 1}^{N_f} \frac{1}{\left(1- z_a {}u_i a x^r\right) \left(1- z_a^{-1} \tilde{{}u}_i a x^r\right)} \\
&\qquad +\sum_{m_{2} = -\infty}^{-1} \tau^{-m_2} a^{-N_f m_2} x^{(1-r) N_f m_2} \oint \ud\mu_{U(N_c-1)} \prod_{a = 1}^{N_c-1} \prod_{i = 1}^{N_f} \frac{1}{\left(1-z_a {}u_i a x^r\right) \left(1-z_a^{-1} \tilde{{}u}_i a x^r\right)} \\
&= g(x,a,{}u,\tilde{{}u};\mathcal{M}^{0}_{U(N_c),N_f})+\frac{\tau^{-1} a^{-N_f} x^{(1-r) N_f-N_c+1}}{1-\tau^{-1} a^{-N_f} x^{(1-r) N_f-N_c+1}} \times g(x,a,{}u,\tilde{{}u};\mathcal{M}^{0}_{U(N_c-1),N_f}) \\
&= g(\mathcal{M}^{0}_{U(N_c),N_f})+g(\mathcal{M}^{-}_{U(N_c),N_f}) -g(\mathcal{M}^{0}_{U(N_c),N_f}\cap \mathcal{M}^{-}_{U(N_c),N_f})
~,~
\end{aligned}
\end{align}
where 
\beal{esx2001}
\mathcal{M}^{0}_{U(N_c),N_f}\cap \mathcal{M}^{-}_{U(N_c),N_f}
=
\mathcal{M}^{0}_{U(N_c-1),N_f} ~.~
\eea

As we observed in section \ref{sec:boundary I/III}, $\mathcal{M}_{U(N_c),N_f}/\left<M_i^j = 0\right>$ is the same as component $\mathcal{M}_{U(N_c),N_f}^{+-}$ for a $U(2)$ theory. Thus, for a $U(2)$ theory, we can completely recover the Hilbert series for each of the four components of the moduli space from those of the four subspaces we have examined.
\\

\paragraph{Superconformal Index and Hilbert Series.} In contrast to the $N_c = 1,2$ cases, the Hilbert series of component $\mathcal{M}_{U(N_c),N_f}^{+-}$ for $N_c \geq 3$ cannot be reproduced as a limit of the superconformal index. Because of this reason, one cannot obtain the exact Hilbert series of a $U(N_c)$ theory with $N_c \geq 3$ by taking a limit of the superconformal index. The index contribution of a chiral ring element in component $\mathcal{M}_{U(N_c),N_f}^{+-}$ could cancel with the contribution of another fermionic operator. In that case any analytic manipulation of the superconformal index, for example taking a limit of the index, cannot trace the contribution of that chiral ring element. Let us consider an example. If we consider the $U(3)$ theory with five flavors, there is a chiral ring element of the form $v_+ v_- M_{(i_1}^{(j_1} M_{i_2}^{j_2} M_{i_3}^{j_3} M_{i_4}^{j_4} M_{i_5)}^{j_5)}$, which has $E+j = 6$ and transforms in the representation $[0,0,0,5] \times [5,0,0,0]$ of $SU(5)_1 \times SU(5)_2$ whose dimension is given by $126^2 = 15876$, and most crucially has charges $A=0$, $T=0$. These charges make it easy to identify many non-zero spin operators. Most of them are fermionic such that their contributions come with a negative sign and could cancel the contributions of $v_+ v_- M_{(i_1}^{(j_1} M_{i_2}^{j_2} M_{i_3}^{j_3} M_{i_4}^{j_4} M_{i_5)}^{j_5)}$. For example, the index contributions of $v_+ v_- M_{(i_1}^{(j_1} M_{i_2}^{j_2} M_{i_3}^{j_3} M_{i_4}^{j_4} M_{i_5)}^{j_5)}$ contain the following terms:
\begin{align} \label{eq:v2M5}
\ldots+u_1^5 x^6+u_1^4 u_2 x^6+\ldots.
\end{align}
On the other hand, the index contributions of the nonzero spin states contain 
\begin{align}
\ldots-u_1^4 u_2 x^6+\ldots,
\end{align}
which comes from $(Q_1 \psi_Q^\dagger{}^3) (Q_1 \psi_Q^\dagger{}^4) (Q_1 \psi_Q^\dagger{}^5)$. That contribution cancels out the term $u_1^4 u_2 x^6$ in \eqref{eq:v2M5}. On the other hand, the other term $u_1^5 x^6$ in \eqref{eq:v2M5} does not appear in the contributions of the nonzero spin states. Therefore, the cancelation of $u_1^4 u_2 x^6$ is accidental. In fact there are many cancelations between the contributions of $v_+ v_- M_{(i_1}^{(j_1} M_{i_2}^{j_2} M_{i_3}^{j_3} M_{i_4}^{j_4} M_{i_5)}^{j_5)}$ and those of the nonzero spin states. Because of these cancellations, taking the limit of the index does not capture the presence of this operator in the chiral ring.
\\

%========================================%
%========================================%

\section*{Acknowledgements}

A. H., J. P.  and  R.-K. S. gratefully acknowledge hospitality at the Simons Center for Geometry and Physics, Stony Brook University where some of the research for this paper was performed.
A. H. is grateful for the hospitality of the Korea Institute for Advanced Study in Seoul and acknowledges private communication and invaluable discussions with Stefano Cremonesi. He is also grateful for discussions with Alberto Zaffaroni and Noppadol Mekareeya.
A. H. and R.-K. S. are grateful for the hospitality of the ICMS in Edinburgh.
H. K. is supported by the NRF-2013-Fostering Core Leaders of the Future Basic Science Program.
J. P. is  supported in part by the National Research Foundation
of Korea Grants No. 2012R1A1A2009117, 2012R1A2A2A06046278. J.P. also appreciates APCTP for its stimulating environment
for research.
\\

%========================================%
%========================================%
\bibliographystyle{JHEP}
\bibliography{mybib}

\providecommand{\href}[2]{#2}\begingroup\raggedright\begin{thebibliography}{10}

\bibitem{Benvenuti:2006qr}
S.~Benvenuti, B.~Feng, A.~Hanany, and Y.-H. He, {\it {Counting BPS operators in
  gauge theories: Quivers, syzygies and plethystics}},  {\em JHEP} {\bf 11}
  (2007) 050, [\href{http://xxx.lanl.gov/abs/hep-th/0608050}{{\tt
  hep-th/0608050}}].

\bibitem{Hanany:2006uc}
A.~Hanany and C.~Romelsberger, {\it {Counting BPS operators in the chiral ring
  of N = 2 supersymmetric gauge theories or N = 2 braine surgery}},  {\em Adv.
  Theor. Math. Phys.} {\bf 11} (2007) 1091--1112,
  [\href{http://xxx.lanl.gov/abs/hep-th/0611346}{{\tt hep-th/0611346}}].

\bibitem{Feng:2007ur}
B.~Feng, A.~Hanany, and Y.-H. He, {\it {Counting Gauge Invariants: the
  Plethystic Program}},  {\em JHEP} {\bf 03} (2007) 090,
  [\href{http://xxx.lanl.gov/abs/hep-th/0701063}{{\tt hep-th/0701063}}].

\bibitem{Butti:2007jv}
A.~Butti, D.~Forcella, A.~Hanany, D.~Vegh, and A.~Zaffaroni, {\it {Counting
  Chiral Operators in Quiver Gauge Theories}},  {\em JHEP} {\bf 11} (2007) 092,
  [\href{http://xxx.lanl.gov/abs/0705.2771}{{\tt arXiv:0705.2771}}].

\bibitem{Hanany:2007zz}
A.~Hanany, {\it {Counting BPS operators in the chiral ring: The plethystic
  story}},  {\em AIP Conf.Proc.} {\bf 939} (2007) 165--175.

\bibitem{Forcella:2008bb}
D.~Forcella, A.~Hanany, Y.-H. He, and A.~Zaffaroni, {\it {The Master Space of
  N=1 Gauge Theories}},  {\em JHEP} {\bf 0808} (2008) 012,
  [\href{http://xxx.lanl.gov/abs/0801.1585}{{\tt arXiv:0801.1585}}].

\bibitem{Forcella:2008eh}
D.~Forcella, A.~Hanany, Y.-H. He, and A.~Zaffaroni, {\it {Mastering the Master
  Space}},  {\em Lett.Math.Phys.} {\bf 85} (2008) 163--171,
  [\href{http://xxx.lanl.gov/abs/0801.3477}{{\tt arXiv:0801.3477}}].

\bibitem{Cremonesi:2013lqa}
S.~Cremonesi, A.~Hanany, and A.~Zaffaroni, {\it {Monopole operators and Hilbert
  series of Coulomb branches of $3d$ $\mathcal{N} = 4$ gauge theories}},  {\em
  JHEP} {\bf 1401} (2014) 005, [\href{http://xxx.lanl.gov/abs/1309.2657}{{\tt
  arXiv:1309.2657}}].

\bibitem{Seiberg:1994pq}
N.~Seiberg, {\it {Electric - magnetic duality in supersymmetric nonAbelian
  gauge theories}},  {\em Nucl.Phys.} {\bf B435} (1995) 129--146,
  [\href{http://xxx.lanl.gov/abs/hep-th/9411149}{{\tt hep-th/9411149}}].

\bibitem{Aharony:1997gp}
O.~Aharony, {\it {IR duality in d = 3 N=2 supersymmetric USp(2N(c)) and U(N(c))
  gauge theories}},  {\em Phys.Lett.} {\bf B404} (1997) 71--76,
  [\href{http://xxx.lanl.gov/abs/hep-th/9703215}{{\tt hep-th/9703215}}].

\bibitem{Aharony:1997bx}
O.~Aharony, A.~Hanany, K.~A. Intriligator, N.~Seiberg, and M.~Strassler, {\it
  {Aspects of N=2 supersymmetric gauge theories in three-dimensions}},  {\em
  Nucl.Phys.} {\bf B499} (1997) 67--99,
  [\href{http://xxx.lanl.gov/abs/hep-th/9703110}{{\tt hep-th/9703110}}].

\bibitem{Karch:1997ux}
A.~Karch, {\it {Seiberg duality in three-dimensions}},  {\em Phys.Lett.} {\bf
  B405} (1997) 79--84, [\href{http://xxx.lanl.gov/abs/hep-th/9703172}{{\tt
  hep-th/9703172}}].

\bibitem{Witten:1999ds}
E.~Witten, {\it {Supersymmetric index of three-dimensional gauge theory}},
  \href{http://xxx.lanl.gov/abs/hep-th/9903005}{{\tt hep-th/9903005}}.

\bibitem{Kapustin:2011vz}
A.~Kapustin, H.~Kim, and J.~Park, {\it {Dualities for 3d Theories with Tensor
  Matter}},  {\em JHEP} {\bf 1112} (2011) 087,
  [\href{http://xxx.lanl.gov/abs/1110.2547}{{\tt arXiv:1110.2547}}].

\bibitem{Kim:2009wb}
S.~Kim, {\it {The Complete superconformal index for N=6 Chern-Simons theory}},
  {\em Nucl.Phys.} {\bf B821} (2009) 241--284,
  [\href{http://xxx.lanl.gov/abs/0903.4172}{{\tt arXiv:0903.4172}}].

\bibitem{Imamura:2011su}
Y.~Imamura and S.~Yokoyama, {\it {Index for three dimensional superconformal
  field theories with general R-charge assignments}},  {\em JHEP} {\bf 1104}
  (2011) 007, [\href{http://xxx.lanl.gov/abs/1101.0557}{{\tt
  arXiv:1101.0557}}].

\bibitem{Bhattacharya:2008bja}
J.~Bhattacharya and S.~Minwalla, {\it {Superconformal Indices for N = 6 Chern
  Simons Theories}},  {\em JHEP} {\bf 0901} (2009) 014,
  [\href{http://xxx.lanl.gov/abs/0806.3251}{{\tt arXiv:0806.3251}}].

\bibitem{Bhattacharya:2008zy}
J.~Bhattacharya, S.~Bhattacharyya, S.~Minwalla, and S.~Raju, {\it {Indices for
  Superconformal Field Theories in 3,5 and 6 Dimensions}},  {\em JHEP} {\bf
  0802} (2008) 064, [\href{http://xxx.lanl.gov/abs/0801.1435}{{\tt
  arXiv:0801.1435}}].

\bibitem{Hwang:2011ht}
C.~Hwang, K.-J. Park, and J.~Park, {\it {Evidence for Aharony duality for
  orthogonal gauge groups}},  {\em JHEP} {\bf 1111} (2011) 011,
  [\href{http://xxx.lanl.gov/abs/1109.2828}{{\tt arXiv:1109.2828}}].

\bibitem{Bashkirov:2011vy}
D.~Bashkirov, {\it {Aharony duality and monopole operators in three
  dimensions}},  \href{http://xxx.lanl.gov/abs/1106.4110}{{\tt
  arXiv:1106.4110}}.

\bibitem{Hwang:2011qt}
C.~Hwang, H.~Kim, K.-J. Park, and J.~Park, {\it {Index computation for 3d
  Chern-Simons matter theory: test of Seiberg-like duality}},  {\em JHEP} {\bf
  1109} (2011) 037, [\href{http://xxx.lanl.gov/abs/1107.4942}{{\tt
  arXiv:1107.4942}}].

\bibitem{Kim:2013cma}
H.~Kim and J.~Park, {\it {Aharony Dualities for 3d Theories with Adjoint
  Matter}},  {\em JHEP} {\bf 1306} (2013) 106,
  [\href{http://xxx.lanl.gov/abs/1302.3645}{{\tt arXiv:1302.3645}}].

\bibitem{Benvenuti:2010pq}
S.~Benvenuti, A.~Hanany, and N.~Mekareeya, {\it {The Hilbert Series of the One
  Instanton Moduli Space}},  {\em JHEP} {\bf 1006} (2010) 100,
  [\href{http://xxx.lanl.gov/abs/1005.3026}{{\tt arXiv:1005.3026}}].

\bibitem{Hanany:2012dm}
A.~Hanany, N.~Mekareeya, and S.~S. Razamat, {\it {Hilbert Series for Moduli
  Spaces of Two Instantons}},  {\em JHEP} {\bf 1301} (2013) 070,
  [\href{http://xxx.lanl.gov/abs/1205.4741}{{\tt arXiv:1205.4741}}].

\bibitem{Dey:2013fea}
A.~Dey, A.~Hanany, N.~Mekareeya, D.~Rodríguez-Gómez, and R.-K. Seong, {\it
  {Hilbert Series for Moduli Spaces of Instantons on
  $\mathbb{C}$$^{2}$/$\mathbb{Z}$$_{n}$}},  {\em JHEP} {\bf 1401} (2014) 182,
  [\href{http://xxx.lanl.gov/abs/1309.0812}{{\tt arXiv:1309.0812}}].

\bibitem{Cremonesi:2014xha}
S.~Cremonesi, G.~Ferlito, A.~Hanany, and N.~Mekareeya, {\it {Coulomb Branch and
  The Moduli Space of Instantons}},
  \href{http://xxx.lanl.gov/abs/1408.6835}{{\tt arXiv:1408.6835}}.

\bibitem{Hanany:2014hia}
A.~Hanany and R.-K. Seong, {\it {Hilbert Series and Moduli Spaces of k U(N)
  Vortices}},  \href{http://xxx.lanl.gov/abs/1403.4950}{{\tt arXiv:1403.4950}}.

\bibitem{Hanany:2012hi}
A.~Hanany and R.-K. Seong, {\it {Brane Tilings and Reflexive Polygons}},  {\em
  Fortsch.Phys.} {\bf 60} (2012) 695--803,
  [\href{http://xxx.lanl.gov/abs/1201.2614}{{\tt arXiv:1201.2614}}].

\bibitem{Hanany:2012vc}
A.~Hanany and R.-K. Seong, {\it {Brane Tilings and Specular Duality}},  {\em
  JHEP} {\bf 1208} (2012) 107, [\href{http://xxx.lanl.gov/abs/1206.2386}{{\tt
  arXiv:1206.2386}}].

\bibitem{Cremonesi:2014kwa}
S.~Cremonesi, A.~Hanany, N.~Mekareeya, and A.~Zaffaroni, {\it {Coulomb branch
  Hilbert series and Hall-Littlewood polynomials}},  {\em JHEP} {\bf 1409}
  (2014) 178, [\href{http://xxx.lanl.gov/abs/1403.0585}{{\tt
  arXiv:1403.0585}}].

\bibitem{Cremonesi:2014vla}
S.~Cremonesi, A.~Hanany, N.~Mekareeya, and A.~Zaffaroni, {\it {Coulomb branch
  Hilbert series and Three Dimensional Sicilian Theories}},  {\em JHEP} {\bf
  1409} (2014) 185, [\href{http://xxx.lanl.gov/abs/1403.2384}{{\tt
  arXiv:1403.2384}}].

\bibitem{Aharony:2013dha}
O.~Aharony, S.~S. Razamat, N.~Seiberg, and B.~Willett, {\it {3d dualities from
  4d dualities}},  {\em JHEP} {\bf 1307} (2013) 149,
  [\href{http://xxx.lanl.gov/abs/1305.3924}{{\tt arXiv:1305.3924}}].

\bibitem{callias1978}
C.~Callias, {\it Axial anomalies and index theorems on open spaces},  {\em
  Comm. Math. Phys.} {\bf 62} (1978), no.~3 213--234.

\bibitem{unpubcremonesi}
S.~Cremonesi, {\it {to be published}}, .

\bibitem{Aharony:1997ju}
O.~Aharony and A.~Hanany, {\it {Branes, superpotentials and superconformal
  fixed points}},  {\em Nucl.Phys.} {\bf B504} (1997) 239--271,
  [\href{http://xxx.lanl.gov/abs/hep-th/9704170}{{\tt hep-th/9704170}}].

\bibitem{'tHooft19781}
G.~'t~Hooft, {\it On the phase transition towards permanent quark confinement},
   {\em Nuclear Physics B} {\bf 138} (1978), no.~1 1 -- 25.

\bibitem{PhysRevD.14.2728}
F.~Englert and P.~Windey, {\it Quantization condition for 't hooft monopoles in
  compact simple lie groups},  {\em Phys. Rev. D} {\bf 14} (Nov, 1976)
  2728--2731.

\bibitem{Goddard19771}
P.~Goddard, J.~Nuyts, and D.~Olive, {\it Gauge theories and magnetic charge},
  {\em Nuclear Physics B} {\bf 125} (1977), no.~1 1 -- 28.

\bibitem{Kapustin:2005py}
A.~Kapustin, {\it {Wilson-'t Hooft operators in four-dimensional gauge theories
  and S-duality}},  {\em Phys.Rev.} {\bf D74} (2006) 025005,
  [\href{http://xxx.lanl.gov/abs/hep-th/0501015}{{\tt hep-th/0501015}}].

\bibitem{Safdi:2012re}
B.~R. Safdi, I.~R. Klebanov, and J.~Lee, {\it {A Crack in the Conformal
  Window}},  {\em JHEP} {\bf 1304} (2013) 165,
  [\href{http://xxx.lanl.gov/abs/1212.4502}{{\tt arXiv:1212.4502}}].

\bibitem{Willett:2011gp}
B.~Willett and I.~Yaakov, {\it {N=2 Dualities and Z Extremization in Three
  Dimensions}},  \href{http://xxx.lanl.gov/abs/1104.0487}{{\tt
  arXiv:1104.0487}}.

\bibitem{Benini:2011mf}
F.~Benini, C.~Closset, and S.~Cremonesi, {\it {Comments on 3d Seiberg-like
  dualities}},  {\em JHEP} {\bf 1110} (2011) 075,
  [\href{http://xxx.lanl.gov/abs/1108.5373}{{\tt arXiv:1108.5373}}].

\bibitem{Hwang:2012jh}
C.~Hwang, H.-C. Kim, and J.~Park, {\it {Factorization of the 3d superconformal
  index}},  {\em JHEP} {\bf 1408} (2014) 018,
  [\href{http://xxx.lanl.gov/abs/1211.6023}{{\tt arXiv:1211.6023}}].

\bibitem{Gray:2008yu}
J.~Gray, A.~Hanany, Y.-H. He, V.~Jejjala, and N.~Mekareeya, {\it {SQCD: A
  Geometric Apercu}},  {\em JHEP} {\bf 0805} (2008) 099,
  [\href{http://xxx.lanl.gov/abs/0803.4257}{{\tt arXiv:0803.4257}}].

\bibitem{Razamat:2014pta}
S.~S. Razamat and B.~Willett, {\it {Down the rabbit hole with theories of class
  $ \mathcal{S} $}},  {\em JHEP} {\bf 1410} (2014) 99,
  [\href{http://xxx.lanl.gov/abs/1403.6107}{{\tt arXiv:1403.6107}}].

\bibitem{2010NuPhB.827..183I}
Y.~{Imamura} and S.~{Yokoyama}, {\it {A monopole index for N=4 Chern-Simons
  theories}},  {\em Nuclear Physics B} {\bf 827} (Mar., 2010) 183--216,
  [\href{http://xxx.lanl.gov/abs/0908.0988}{{\tt arXiv:0908.0988}}].

\end{thebibliography}\endgroup

%========================================%

\end{document}